\documentclass[aps,pra,reprint,amsmath,amssymb,superscriptaddress,showkeys]{revtex4-2}
\usepackage{bm,psfrag,setspace,bbold,braket}
\usepackage{mathrsfs}
\usepackage{amsthm}
\usepackage{braket}
\usepackage{hhline}
\usepackage[T1]{fontenc}
\usepackage[left]{lineno}
\usepackage[pdftex]{graphicx}
\usepackage{accents}
\usepackage{hyperref}
\usepackage[dvipsnames]{xcolor}
\hypersetup{colorlinks=true,linkcolor=RoyalBlue,citecolor=RoyalBlue,urlcolor=RoyalBlue,hypertexnames=true}
\selectfont

\usepackage{orcidlink}

\begin{document}
\setlength{\parskip}{0pt}
\title{From Wavefunctional Entanglement to Entangled Wavefunctional Degrees of Freedom}
\author{Aniruddha Bhattacharya\,\orcidlink{0000-0002-7727-0514}}
\email{Contact author: anirudb@umich.edu, he/him/his}
\affiliation{The School of Physics, The Georgia Institute of Technology, 837 State Street NW, Atlanta, Georgia 30332-0430, USA}
\date{\today}

\begin{abstract}

The question of whether entanglement between photons is equivalent to entanglement between their characteristic field modes—specifically, the single-particle wavefunctions that are composed and superposed to describe particles in such modes—is a key, open problem concerning multi-partite optical degrees of freedom, and has profound implications for topics ranging from quantum foundations to quantum computation. Here, I offer a fresh, deeper, physical insight into this subtle, albeit enduring, issue by describing a situation in which entangling interactions between optical modes—namely, the wavefunctions—can be distilled into genuine entanglement between the physical, observable properties of the photons—which are the wavefunctional degrees of freedom. This theoretical observation also highlights the salience of the measurement context—especially, of clearly disambiguating between the choice of the quantum subsystem and the decision to measure an observable along a particular axis of measurement—while quantifying and transforming quantum optical entanglement. \textcolor{RoyalBlue}{This theoretical observation might be applied to formulate a new class of protocols for performing quantum information tasks, using entangled photons within inseparable field modes.}

\end{abstract}

\maketitle

\section{Introduction}\label{sec:Introduce}

Quantum entanglement and quantum nonlocality \cite{PhysRev.47.777, PhysRev.48.696, schrodinger1935gegenwartige, Schrödinger_1935, bell2004speakable, PhysRevLett.71.1665, popescu1994quantum, RevModPhys.81.865, RevModPhys.81.1727, RevModPhys.85.1103, RevModPhys.86.419} are fundamental features of the quantum mechanical description of many-particle systems, and play defining roles in foundations of quantum physics \cite{RevModPhys.71.S288, bertlmann2013quantum, bertlmann2013quantum2, RevModPhys.92.021002}; quantum information science and engineering \cite{Bouwmeester2000, doi:10.1098/rspa.2002.1097, nielsen2010quantum, preskill2012quantum}; atomic \cite{RevModPhys.73.565, evered2023high, joshi2023exploring, iqbal2024non, PhysRevLett.132.113601, grinkemeyer2025error}, molecular \cite{bao2023dipolar}, optical \cite{PhysRevLett.28.938, PhysRevLett.47.460, PhysRevLett.49.91, PhysRevLett.49.1804, PhysRevLett.81.5039, RevModPhys.84.777}, condensed matter \cite{LAFLORENCIE20161, PhysRevLett.105.151602}, and high-energy physics \cite{RevModPhys.90.045003}; as well as in cosmology \cite{PhysRevLett.96.181602, https://doi.org/10.1002/prop.201300020, penington2020entanglement, pg4r-fy8n}. While the earliest intimations of the notion of quantum nonlocality are traceable to the 1927 Solvay Conference \cite{*[{See, for example, }] [{ for a history of thought on entanglement, nonlocality, and incompleteness.}] Bacciagaluppi_Crull_2024, PhysRevLett.73.2279, bacciagaluppi2009quantum}, in 1935, Einstein, Podolsky, and Rosen (E.P.R.) \cite{PhysRev.47.777} formally introduced the concept of quantum entanglement in the course of devising a \emph{Gedankenexperiment} to clearly show a conflict between: $(1)$ the quantum mechanical postulate that the wavefunction provides a complete description of the physical reality of a quantum system; and $(2)$ the absence of simultaneous elements of physical reality corresponding to values of mutually complementary physical quantities that are described by non-commuting operators. Specifically, E.P.R. argued that either $(1)$ or $(2)$ is true, but not both. While, in the standard quantum theory, $(1)$ is true and $(2)$ is false, E.P.R. were able to demonstrate that the situation in which the coordinates, as well as the momenta of two particles are entangled, paradoxically suggests that $(1)$ is false, but $(2)$ is true.

A more careful reading of their paper reveals that E.P.R. devised entanglement as a mechanism to predict the value of a physical quantity of a quantum system, say, system I with \emph{certainty} \cite{PhysRev.47.777}—that is, with probability equal to unity, like in a classical, realist theory—without in any way \emph{disturbing} \cite{PhysRev.47.777}—that is, altering the physical state of—the system. The actual, direct measurement were to be carried out on the system's entangled, spatially separated counterpart, say, system II. Crucially, E.P.R. implicitly assumed that the principle of classical locality should ensure that such an indirect measurement on system II would not \emph{disturb} system I \cite{PhysRev.48.696, Schilpp1949-SCHTLO-38}.

Guided by Bohm's reformulation of the E.P.R. paradox \cite{PhysRev.108.1070} in terms of discrete-valued, dichotomous observables, namely, spin angular momenta, Bell correctly realized that the above-described conflict is actually one between the irrevocable indeterminism of quantum mechanics and E.P.R.'s presupposed, cherished notion of local, classical realism \cite{PhysicsPhysiqueFizika.1.195, *bell2004speakablepaper2}. Simply put, situations, such as the E.P.R.B. (E.P.R.-Bohm) entanglement in which there are perfect correlations—or anti-correlations—between the outcomes of measurements performed upon two causally disconnected, space-like separated particles suggest that there are two opposing descriptions of quantum reality, namely, either: $(3)$ physical reality is nonlocal, measurement outcomes are inescapably probabilistic, and the \textcolor{RoyalBlue}{wavefunctional} description of physical reality is complete; or $(4)$ physical reality is local, measurement outcomes are predetermined by hidden elements of reality, and, therefore, the \textcolor{RoyalBlue}{wavefunctional} description of physical reality is incomplete, which is the crux of the E.P.R. argument \cite{maudlin2014bell, *werner2014comment, *wiseman2014two}.

Notably, Bell resolved the E.P.R. paradox by formulating empirically verifiable inequalities \cite{PhysicsPhysiqueFizika.1.195, *bell2004speakablepaper2} that quantified the difference between $(3)$ and $(4)$ above. Bell's inequalities bound from above the strengths of correlations between the outcomes of measurements on two quantum systems in situations in which each of such systems has its own, individual properties, as is suggested by local realist theories that are formulated in the spirit of the E.P.R. viewpoint, like $(4)$ above.

A series of exceedingly elegant and increasingly loophole-free experiments, which were carried out between 1972 and the present time—see, for example, Refs. \cite{PhysRevLett.28.938, PhysRevLett.47.460, PhysRevLett.49.91, PhysRevLett.49.1804, PhysRevLett.81.5039, PhysRevLett.115.250401, PhysRevLett.115.250402, hensen2015loophole, aspect2015closing, PhysRevLett.118.060401, yin2017satellite}, and references therein—conclusively demonstrated an explicit violation of refined versions of Bell's inequalities \cite{PhysRevLett.23.880, PhysRevD.10.526, peres1999all}, such as the Bell-C.H.S.H. inequalities \cite{PhysRevLett.23.880}. These experiments are enormously significant, as they decisively rule out any conceivable, deterministic, local realist, hidden-variables theory \cite{RevModPhys.38.447, *bell2004speakablepaper1} that attribute the statistical nature of measurement outcomes to the underlying statistical distributions of hidden, supplementary elements of physical reality, analogous to classical probability distributions of classical random variables.

Strikingly, Greenberger, Horne, and Zeilinger (G.H.Z.) extended Bell's results by showing that—under special circumstances—the definite, certain predictions about measurement outcomes by standard quantum theory directly conflict the corresponding predictions by local realist theories, as opposed to a mere statistical contradiction involving inequalities based on correlation functions \cite{greenberger1989going, greenberger1990bell}. Such situations are realized by probing specific kinds of multi-particle entangled states, known as the G.H.Z. states \cite{PhysRevLett.82.1345, PhysRevLett.86.4435, RevModPhys.84.777}. $N$-particle G.H.Z. states, where $N \geq 3$, are commonly used in optical quantum computing schemes and protocols \cite{walther2004broglie, PhysRevLett.93.020504, PhysRevLett.92.017902, PhysRevLett.94.070402, PhysRevLett.94.240501, PhysRevLett.97.020501, walther2005experimental, prevedel2007high, PhysRevLett.99.250503, PhysRevLett.98.140501} that, fundamentally, derive their quantum advantage from the intrinsic quantum nonlocality of such highly entangled states \cite{pan2000experimental, walther2004broglie}. More broadly, multi-particle quantum entanglement and quantum contextuality \cite{RevModPhys.38.447, *bell2004speakablepaper1, kochenspecker1967, howard2014contextuality}—which is a generalized notion of quantum nonlocality and plays an essential role, for example, in magic state distillation protocols \cite{gupta2024msd, lukin2025msd}—have been collectively suggested \cite{howard2014contextuality} as fundamental sources of scalable quantum computational advantage \cite{bluvstein2022quantum, evered2023high, bluvstein2024logical} over their classical counterparts. Moreover, the quantum mechanical predictions for these multi-particle entangled states are completely independent of the relative arrangements of the measuring apparatuses in space and time.

\begin{figure*}[ht]
\includegraphics[width=\textwidth]{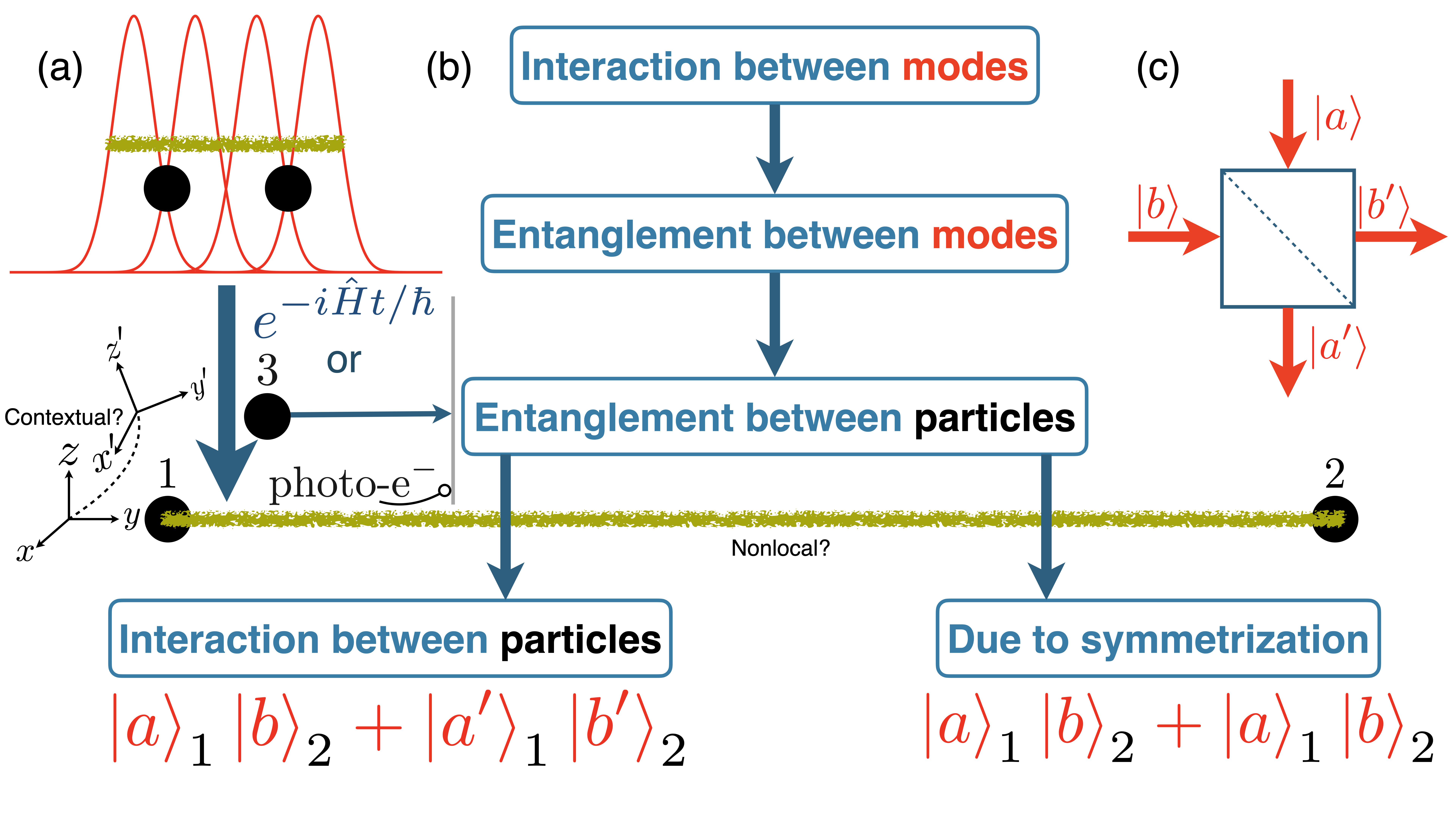}
\caption{\label{fig:f1} \textbf{The problem of transforming entanglement between modes to useful entanglement between particles.} (a) This paper raises and addresses the question of how entanglement between optical field modes (shown in red) can be—either, possibly, unitarily, or by making a measurement on an ancillary photon—distilled into entanglement between the photons (depicted in black) that are described by such modes. (b) A flow chart sequentially deconstructing this problem from its simplest—entangling modes—to its deepest—creating genuinely entangled states of interacting photons (see bottom left), which are useful for quantum computation, as opposed to states of non-interacting photons that are entangled merely due to symmetrization (see bottom right)—levels. (c) An interferometer—such as a beam splitter—should be able to provision the modes for constructing the states described in (b). \textcolor{RoyalBlue}{Notice that this highly simplified situation is shown merely as an illustrative example; in practice, however, the two input and the two output modes are rarely useful for creating entangled states of two photons and four modes.}}
\end{figure*}

The quantum mechanical systems of choice in a majority of the above-described experiments were optical, which raised an accompanying, profound question: (Qu1) Is entanglement between modes of light equivalent to entanglement between individual particles of light? Resolution of this quantum foundational issue also has significant implications for quantum information processing \cite{PhysRevLett.120.260502} and quantum-enhanced sensing \cite{PhysRevLett.132.190001} with highly entangled states of light. \textcolor{RoyalBlue}{While Refs. \cite{dalton2017quantum, dalton2017quantum2, PhysRevLett.134.080201} comprehensively survey the relevant theoretical and experimental advancements in understanding the similarities and the differences between mode-mode entanglement and particle-particle entanglement, the remainder of this introduction offers a selection of the most illuminating and interesting aspects—of such advances—so as to motivate the importance of elucidating whether entanglement between optical photons is equivalent to entanglement between their characteristic optical field modes.}

\textcolor{RoyalBlue}{The following discussion is inspired, in particular, by the systematic survey offered in the End Matter of Ref. \cite{PhysRevLett.134.080201}. As is well known and is well understood, it is essential to clearly and unambiguously identify the relevant quantum subsystems—as well as the salient quantum observables that are entangled across such quantum subsystems—so as to properly and rigorously characterize and quantify quantum mechanical entanglement. For the case of composite, multi-partite, quantum optical systems, one can consider either the modes of light, or the particles of light, as the individual quantum subsystems. In the literature, usually, distinguishable modes are regarded as the more suitable candidate, since such modes have all the essential characteristics of well-defined quantum subsystems \cite{dalton2017quantum,10.1093/acprof:oso/9780199215706.001.0001,PhysRevA.66.052323,PhysRevA.67.013609,cunha2007entanglement,BENATTI2010924,Benatti_2011,walther2004broglie,Dalton_2014}. For instance, modes can be distinguishable; modes are directly, physically accessible and are amenable to measurements \cite{RevModPhys.81.299}; and, finally, modes can exist as separate, spatially distinct systems that can be prepared in quantum states that correspond to that individual system alone. In contrast, considering mathematically labeled, individual, indistinguishable particles as quantum subsystems appears to be more of a mathematical construct, as compared with a physically reasonable approach, because such particles do not fulfill the above-described criteria, for well-defined quantum subsystems. Specifically, Peres has emphasized that the properly exchange-symmetrized wavefunctions of two identical particles are always entangled, if one views the individual, indistinguishable particles as the subsystems. This notion is often referred to as \emph{entanglement due to symmetrization} \cite{peres2002quantum}. This entanglement due to symmetrization, however, is usually neither useful nor utilizable for applications in quantum information science and engineering, with the possible exception of the scheme presented in Ref. \cite{PhysRevLett.112.150501}; see, also, the discussion about this issue in Sec. \ref{sec:Equivalence} and Fig. \ref{fig:f2}.}

\textcolor{RoyalBlue}{In general, however, algorithms, protocols, and schemes, in quantum information science and engineering, are designed and implemented, such that they operate with and act upon quantum subsystems, for instance, qubits that are encoded from quantum mechanical particles, such as neutral atoms, trapped ions, atomic nuclei, polar molecules, as well as individual photons and electrons. The present work is, therefore, motivated by the reasonable hope and goal of identifying and constructing genuinely entangled states of optical modes, as well as genuinely entangled states of photons—that circumvents the issue of entanglement due to symmetrization—and ascertaining if and how the former can be transformed into the latter.}

Usually, in quantum optics, a bosonic mode is modeled as a single-particle wavefunction, either in real space, or in some other space, such as momentum space \cite{PhysRevLett.119.173202}. This description, however, does not lend itself to direct and unequivocal correspondence with observables, as opposed to a description based on photons, which are fundamental particles, and, therefore, have unambiguously real properties. For example, a beam of light can be described by multiple equivalent—and all mathematically legitimate—bases of modes, even though one, or a preferred set of such mode bases might be particularly convenient for describing and calculating the properties of the light. This feature leads to the question of the relative nature of entanglement between modes, namely: (Qu2) If one changes to a new mode basis, does the inter-modal entanglement still exist? In contrast, inter-particle entanglement is invariant under basis transformations.

As is well known, protocols for deterministically preparing—that is, without requiring any quantum measurements at any step—entangled photonic states—such as the state, $\alpha\left|a\right>_1\left|b\right>_2+\beta\left|c\right>_1\left|d\right>_2$, where $\left|a\right>$ and $\left|c\right>$ represent two spatial modes of photon 1 (quantum subsystem 1), \textcolor{RoyalBlue}{$\left|b\right>$ and $\left|d\right>$ are two spatial modes of photon 2} (quantum subsystem 2), and $\alpha$ and $\beta$ are suitable normalizing, expansion constants—require either direct or effective interactions between photons, both of which are unusually difficult to realize in practice. On the one hand, coherent, entangled superpositions of light field modes are readily prepared by amplitude and beam splitting, as well as by nonlinear optical processes \cite{merzbacher1998quantumchap23}; recently, for example, an expanded set of such superpositions has been shown to be deterministically realizable by higher-dimensional, noiseless, quantum holonomic approaches \cite{PhysRevLett.134.080201}. On the other hand, the pioneering experiments of Zeilinger and his colleagues have shown that quantum measurements play an important role in the creation of G.H.Z. states; for example, the detection of an ancillary, trigger photon heralds the formation of entanglement between $N \geq 3$ photons.

Consequently, one might pose the question, which seems both fundamentally, as well as practically important: (Qu3) Is it possible to transfer the inter-modal entanglement—which is relatively simpler to achieve—to inter-particle entanglement—which is particularly challenging to realize—as deterministically as possible? The purpose of this paper is to show that, indeed, wavefunctional \footnote{Throughout this paper, the term wavefunctional is used as the adjectival form of wavefunction, and is unrelated to a \emph{wave functional} that plays the role of a wavefunction in a quantized theory of wave fields \cite{Schiff_1955}.} entanglement—that is, entanglement between mathematically inseparable quantum wavefunctions or optical field modes—can be transformed into direct entangling interactions between optical photons that are described by such modes. An additional step that entails making a measurement on an ancillary photon appears to be irreplaceable. Schematically, Fig. \ref{fig:f1} summarizes the problem that this paper addresses.

It is important to note that pairs of indistinguishable optical photons—whose angles of linear polarization are entangled—are routinely produced by spontaneous parametric down-conversion (SPDC)-based setups \cite{PhysRevLett.75.4337}. However, although these sources of entangled photon pairs can be, for example, cascaded \cite{PhysRevA.57.2076, Hannes2010} to generate three-photon entangled states \cite{Hannes2010, Shalm2013}, such highly optically non-linear procedures require complex experimental arrangements \cite{Zhang:16, PhysRevLett.121.250505}, and are plagued by low conversion efficiencies, thereby imposing severe requirement on the pump laser. In this paper, I propose an alternative strategy that entails creating mode-mode entangled states—where the modes contain up to $N \geq 3$ photons—and converting such mode-mode entangled states to genuine photon-photon entangled states. \textcolor{RoyalBlue}{Therefore, when viewed from an alternative perspective, the aim of this work can also be construed as suggesting experimentally feasible ways of transforming the entanglement between two arbitrary, bosonic, optical modes into entanglement between their constituent, characteristic, spatial modes.}

\textcolor{RoyalBlue}{I will now briefly describe how the remainder of this paper is arranged. Section \ref{sec:Disambiguate} highlights the importance of disambiguating between the choices of the quantum subsystems and the pertinent quantum observables, while considering quantum states of entangled photons, and of optical modes that are in photon occupation number entangled relationships with each other. In Sec. \ref{sec:Distinguish}, I describe and distinguish between the two fundamental forms of quantum optical entanglement, namely, wavefunctional entanglement and entangled wavefunctional degrees of freedom. While entanglement between optical modes is a particular case of the former, entanglement between particles of light is a specialized case of the latter. This section also explicitly defines the term, \emph{particles} in this context.}

\textcolor{RoyalBlue}{In Sec. \ref{sec:Equivalence}, an enduring, foundational question in quantum optics, specifically, whether entanglement between optical photons—namely, the particles of light—is equivalent to entanglement between their characteristic field modes, is considered and investigated. In this section, I approach this problem, in a novel manner, by describing a situation, specifically, a \emph{Gedankenexperiment} in which the entangling interactions between optical modes are distilled into genuine entanglement between the wavefunctional—that is, physical—degrees of freedom of the photons. This distillation is accomplished by making measurements on an ancillary photon. This measurement process—specifically, the detection of the ancillary photon—controls an anharmonic potential that is essential for creating the final entangled state of photons. This \emph{Gedankenexperiment} can also be construed as an entangling scheme or protocol. Whereas, in Sec. \ref{sec:Swap}, I compare and contrast the above scheme with standard entanglement swapping procedures, in Sec. \ref{sec:Photodetect}, I describe several consequences of non-ideal photo-detection efficiencies, and how such limitations might be overcome. In Sec. \ref{sec:Qpm}, the above \emph{Gedankenexperiment} is simplified and is analyzed, using the notion of quantum pre-measurement. Finally, in Sec. \ref{sec:Conclude}, I summarize the main conclusions of this paper, and suggest future avenues of fundamental research based on this work.}

\section{Disambiguating between choices of subsystems and of observables}\label{sec:Disambiguate}

Before tackling question (Qu3), let us consider experiments of the kind that were pioneered by Aspect \emph{et al.} \cite{PhysRevLett.47.460, PhysRevLett.49.1804, PhysRevLett.49.91} for demonstrating violations of Bell's inequalities, so as to illustrate the role played by the relative context of the quantum measuring arrangements—specifically, that of local basis transformations on individual subsystems—in determining the specific definitions—or identities—of subsystems and structures of observables.

In such experiments, there is a source \emph{Q}, which emits a pair of entangled photons, whose angles of linear polarization are nonlocally correlated. The two photons propagate in opposite directions and each individually encounters an analyzer of linear polarization—for instance, a linearly polarizing, beam-splitting cube, made of a birefringent crystal—that perfectly transmits the parallel—along the angle of orientation of the analyzer, such as $\theta_{\textrm{A}}$ and $\theta_{\textrm{B}}$—as well as the perpendicular, linearly polarized components along two distinct directions. Each of such optical paths ends with a single-photon detector, thereby correlating the position of the photon with its angle of linear polarization, just like a Stern-Gerlach magnet oriented in space correlates the position of the electron—upon a detector screen—with the component of the spin angular momentum along the direction of the magnetic field. The locations of measurements of these angle of linear polarization—which are the entangled physical quantities—of the photons, $\textrm{A}$ and $\textrm{B}$ are assumed to be separated by a space-like interval.

Suppose that \emph{Q} emits the pure, nonfactorable, E.P.R. state: 
\begin{equation}
\begin{split}
\left|\Psi\left(\nu_{\textrm{A}}, \nu_{\textrm{B}}\right)\right> = \frac{1}{\sqrt{2}}\Bigl\{\left|\textrm{H}\right>_{\textrm{A}}\left|\textrm{H}\right>_{\textrm{B}}+\left|\textrm{V}\right>_{\textrm{A}}\left|\textrm{V}\right>_{\textrm{B}}\Bigr\},
\end{split}\label{eq:Entangled_EPR_State}
\end{equation}
where $\textrm{A}$ and $\textrm{B}$ represent the two quantum subsystems—which are the two distinguishable photons—and $\textrm{H}$ and $\textrm{V}$ represent the horizontal $\left(\theta_{\textrm{A}} = \theta_{\textrm{B}} = 0\right)$ and vertical $\left(\theta_{\textrm{A}} = \theta_{\textrm{B}} = \pi/2\right)$ orientations of the linear polarization, respectively. As is described in Appendix \ref{sec:Context} (see also Fig. \ref{fig:f5}), the above two-particle entangled state can be unitarily transformed to the one below, by performing individual basis transformations upon each of the two subsystems: 
\begin{equation}
\begin{split}
\left|\widetilde{\Psi}\left(\nu_{\textrm{A}}, \nu_{\textrm{B}}\right)\right> &= \cos\left(\theta_{\textrm{A}}-\theta_{\textrm{B}}\right)\Biggl\{\frac{\left|+\right>_{\textrm{A}}\left|+\right>_{\textrm{B}}+\left|-\right>_{\textrm{A}}\left|-\right>_{\textrm{B}}}{\sqrt{2}}\Biggr\}\\
&+ \sin\left(\theta_{\textrm{A}}-\theta_{\textrm{B}}\right)\Biggl\{\frac{\left|+\right>_{\textrm{A}}\left|-\right>_{\textrm{B}}-\left|-\right>_{\textrm{A}}\left|+\right>_{\textrm{B}}}{\sqrt{2}}\Biggr\},
\end{split}\label{eq:Transformed_Entangled_State}
\end{equation}
where the single-particle states, $\left|+\right>$ and $\left|-\right>$ are obtained by unitarily transforming $\left|\textrm{H}\right>$ and $\left|\textrm{V}\right>$, respectively, for a given orientation angle. Performing distinct, local, unitary transformations—albeit of the same kind, namely, rotations by different angles—on the two causally disconnected subsystems is valid, since $\left|\widetilde{\Psi}\left(\nu_{\textrm{A}}, \nu_{\textrm{B}}\right)\right>$ can be used to recover all the well-known expressions for the single- and two-particle, detection probabilities; the coefficients of correlation of linear polarization; as well as the Bell-C.H.S.H. inequalities. The entropies of entanglement of $\left|\Psi\left(\nu_{\textrm{A}}, \nu_{\textrm{B}}\right)\right>$ and $\left|\widetilde{\Psi}\left(\nu_{\textrm{A}}, \nu_{\textrm{B}}\right)\right>$ are identical, and constant, regardless of the relative orientation, $\theta = \theta_{\textrm{A}}-\theta_{\textrm{B}}$. A discussion of the differences between these two representations of the two-particle, entangled state, from a physical point of view, is given in Appendix \ref{sec:Interpret}.

Remarkably, a similar situation can be orchestrated for entanglement between modes that describe neutral, bosonic atoms; the entangled property is the linear momentum of such atoms \cite{PhysRevLett.119.173202}. Four output modes are interfered two-by-two in two distinct spatial locations—with individually controllable phases, $\vartheta_{\textrm{A}}$ and $\vartheta_{\textrm{B}}$—so as to demonstrate Bell nonlocality. The entropies of entanglement are also independent of the relative output phase, $\vartheta = \left(\vartheta_{\textrm{A}}-\vartheta_{\textrm{B}}\right)/2$, which plays the role of $\theta$ above. A key assumption of this work is optical modes can be treated like any other bosonic mode, for example, describing cold atoms.

Intriguingly, for modes, which are in photon occupation number entangled relationships with each other, we encounter this issue: The decision to measure the entangled property along, as it were, a particular axis of measurement—or, more precisely, a specifically chosen basis—and the identities—or the definitions—of the individual quantum subsystems, upon which such quantum measurements will be made, are inextricably linked, even though the Bell’s inequalities have been properly reframed for dealing with situations involving continuous, external degrees of freedom \cite{PhysRevLett.62.2209, PhysRevLett.84.2722, PhysRevA.67.012105, PhysRevLett.93.020401, PhysRevA.84.022105, PhysRevLett.119.173202}. For example, basis transformations concomitantly alter the occupation number—the entangled property—and the definition of the modal creation operator—the identity of the subsystem. For the above-described case of atoms with interfering modes, an additional element of subtlety is introduced by linear momenta—unlike spin angular momenta—along distinct axes commuting with each other. I invite the specialist reader to peruse Appendix \ref{sec:Context} for a deeper dive into this issue, and for a discussion of how such ambiguities could be resolved by interferometry.

\section{Distinguishing between Wavefunctional Entanglement and Entangled Wavefunctional Degrees of Freedom}\label{sec:Distinguish}

\textcolor{RoyalBlue}{The purpose of this section is to introduce and clearly define two key terms, namely, wavefunctional entanglement, and entangled wavefunctional degrees of freedom. One of the central purposes of this paper is to effectively distinguish between these two fundamental forms of quantum entanglement, especially in the context of quantum optical systems and observables. It is important that these two concepts are adequately and sharply defined, as they will play essential, defining roles in the main results of this paper.}

\textcolor{RoyalBlue}{The term, \emph{wavefunctional entanglement} conveys a form of quantum entanglement that is directly evident in, and arises from, the mathematical representation of many-particle wavefunctions as non-factorable, or inseparable, combinations of their constituent, single-particle wavefunctions. In particular, conventional, \emph{mode-mode entanglement}—for instance, entanglement between bosonic, optical, field modes—is a specific kind of wavefunctional entanglement, since, typically, these modes are modeled using single-particle wavefunctions that are suitably composed and superposed, so as to describe particles within such modes. Wavefunctional entanglement, consequently, is a more widely generalizable notion that attempts to indicate a fundamental form of quantum entanglement that originates from the mathematical inseparability of constituent, single-particle wavefunctions.}

\textcolor{RoyalBlue}{The term, \emph{entangled wavefunctional degrees of freedom}, in contrast, signifies genuine, non-local—and, possibly, in certain kinds of situations involving multiple commuting observables, contextual—entanglement between the physical properties of two or more particles. These wavefunctional degrees of freedom are nothing but the physical degrees of freedom that are intrinsic to the particle, such as the photon, and that are fully specified by the wavefunctions, or, more specifically, the state vectors of the particles, together with the appropriate Hermitian operators for the observables. Said differently, these entangled wavefunctional degrees of freedom are characteristic of genuinely entangled quantum states of particles, or of \emph{particle-particle entanglement}. Importantly, the meaning and interpretation of the term, \emph{particles}, which plays a central role in characterizing such entangled wavefunctional degrees of freedom, will depend on the measurement context and the specific details of the entangling protocol. See, for example, the end of Sec. \ref{sec:Equivalence} for a discussion of the meaning of particles in the context of the entangling scheme that is introduced in that section. Simply put, whereas mode-mode entanglement signifies entanglement between wave-like quantum subsystems, with continuous-valued quantum observables, particle-particle entanglement describes entanglement between particle-like quantum subsystems, admitting of discrete-valued quantum observables.}

\textcolor{RoyalBlue}{In particular, with regard to mode-mode entanglement, in this paper, I will specifically consider entanglement between spatial optical modes. To develop a deeper physical intuition into the relationship between mode-mode entanglement and particle-particle entanglement, one might imagine the former as being analogous to entanglement between two distinct, separated regions of space, whereas the latter can be thought of as entanglement between particles in such spaces. Spatial delocalization of the particles of light between the two quantum subsystems—namely, the two spatial optical modes, or the two regions of space—appears to be essential for the equivalence between mode-mode entanglement and particle-particle entanglement.}

\begin{figure*}[ht]
\includegraphics[width=\textwidth]{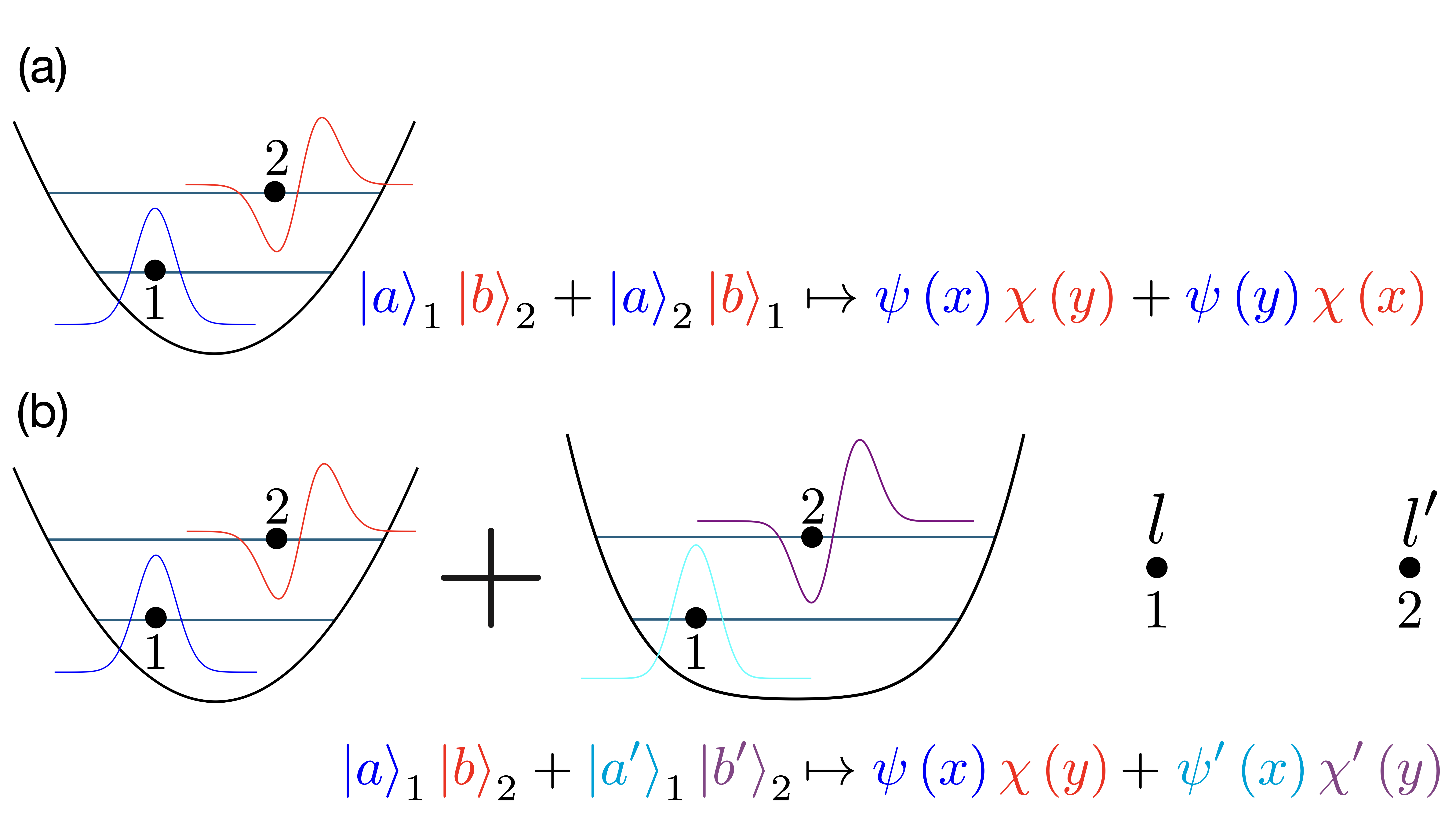}
\caption{\label{fig:f2} \textbf{Comparing and contrasting entanglement between indistinguishable photons, with entanglement between extrinsically distinguishable, albeit intrinsically identical, photons.} (a) The phenomenon of \emph{entanglement due to symmetrization} between two photons that arises as a consequence of the two photons' being identical and conforming to exchange symmetry. The spatial, optical modes of the two photons have been mapped onto the bound energy eigenmodes of a simple harmonic potential, so as to illustrate this situation. (b) Conversely, this figure describes genuine entanglement between two photons that are assumed to distinguishable, due to an extrinsic degree of freedom, such as a suitable component of the orbital angular momentum, which takes the values, $l$ and $l^{\prime}$ for photons 1 and 2, respectively. In such a situation, constructing genuinely entangled states of two photons, as indicated in this figure, requires accessing four spatial, optical modes—of which, $\left|a\right>$ and $\left|b\right>$ are mapped onto the bound energy eigenmodes of an harmonic potential, and $\left|a^{\prime}\right>$ and $\left|b^{\prime}\right>$ are mapped onto the bound energy eigenmodes of an anharmonic potential. See, for example, Sec. \ref{sec:Equivalence} for a discussion of a scheme that produces such entangled states.}
\end{figure*}

\section{The Equivalence between Wavefunctional and Inter-particle Entanglement}\label{sec:Equivalence}

With the above caveats and definitions in mind, we are now ready to address (Qu3). \textcolor{RoyalBlue}{The principal findings, which constitute the central, original results of this paper, are summarized in the following principle:}

\begin{figure*}[ht]
\includegraphics[width=\textwidth]{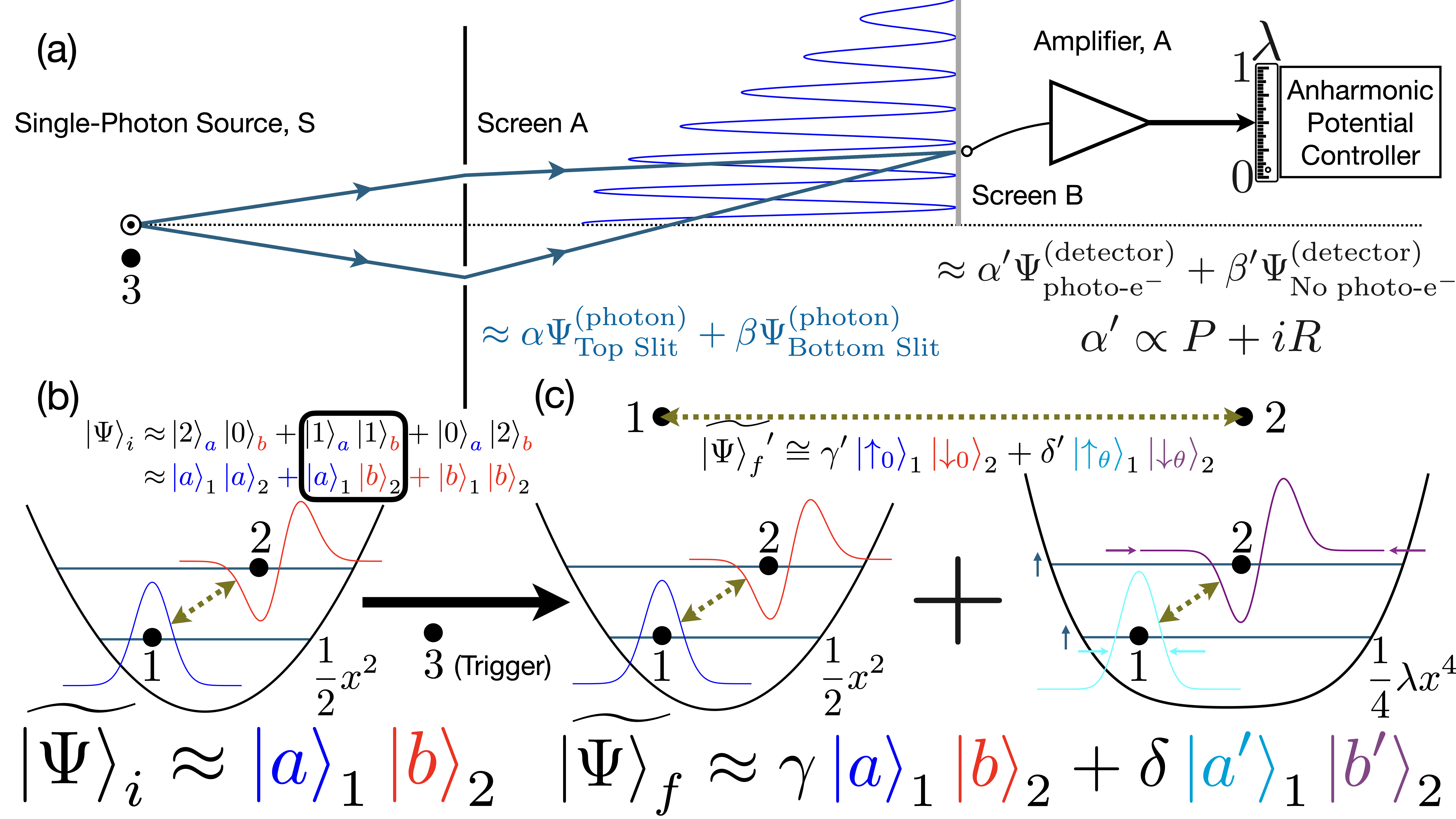}
\caption{\label{fig:f3} \textbf{A simple \emph{Gedankenexperiment} illustrating how entanglement between optical field modes—namely, wavefunctional entanglement—can be distilled into entanglement between physical properties of photons that are described by such modes—namely, the wavefunctional degrees of freedom.} (a) A version of the two-slit diffraction experiment in which the superposition of an ancillary photon (labeled as 3)—of having traveled through the top and bottom slits—is amplified and transformed into a superposition of the photo-detecting apparatus—of having detected and not detected the photon—which, in turn, controls the anharmonicity of an initially harmonic potential, $x^2/2$. The bottom half of the detector screen has been discarded, so as to highlight that, unlike in the corresponding classical case, the location of detection of the particles—for example, the upper half of the screen—is not correlated with the slit of entry, for example, the top slit. (b) The encoding of $\widetilde{\left|\Psi\right>_{i}}$—that describes a pair of modes in a photon number entangled relationship with each other—onto the harmonic oscillator energy eigenmodes (shown in blue and red). (c) Due to the probabilistic nature of the photo-detection of the ancillary photon, $\widetilde{\left|\Psi\right>_{i}}$ transforms into $\widetilde{\left|\Psi\right>_{f}}$, a superposition of a state encoded onto the harmonic oscillator eigenmodes, as well as onto the anharmonic oscillator eigenmodes (shown in cyan and purple). This state is now a genuinely entangled state of two interacting photons (depicted in black, and labeled as 1 and 2), as emphasized by the depiction of an analogous spinor representation in the inset. The vertical and the horizontal lines—alongside the anharmonic potential, $\lambda x^4/4$—represent the increase in the eigenenergies and the spatial narrowing of the eigenfunctions—due to the anharmonicity—respectively.}
\end{figure*}

\textbf{The Equivalence between Wavefunctional and Inter-particle Entanglement Principle.} For a quantum optical system having $M \geq 4$ spatial, optical modes, and operating with $N \geq 2$ input, unentangled photons, the wavefunctional entanglement—that is, the entanglement between the single-particle quantum wavefunctions, or modes—can be transformed into entangling interactions between the photons that are described by such modes.

While I have given a somewhat more formal proof in Appendix \ref{sec:Derive}, \textcolor{RoyalBlue}{as well as a somewhat more rigorous analysis in Sec. \ref{sec:Qpm}}, here, I will outline a simple argument that provides a deeper physical intuition into attacking this formidable problem. Figure \ref{fig:f3} schematically summarizes this physical argument. Inspired by Ref. \cite{PhysRevLett.119.173202}, I will model these spatial, bosonic modes as bound states, as opposed to, for example, propagating Gaussian wavepackets. I am specifically assuming that such modes can be mapped onto the bound energy eigenmodes of a simple harmonic oscillator, and of its anharmonically perturbed analogues; a quantum mechanical particle can always be, meaningfully, said to be placed in a bound state.

Consider, for example, the following input state that describes two modes, which are in a photon occupation number entangled relationship with each other:
\begin{equation}
\begin{split}
\left|\Psi\right>_{i} &= \left|2\right>_{\textrm{a}}\left|0\right>_{\textrm{b}} + \left|1\right>_{\textrm{a}}\left|1\right>_{\textrm{b}}+ \left|0\right>_{\textrm{a}}\left|2\right>_{\textrm{b}}\\
&\equiv \left|a\right>_{\textrm{1}}\left|a\right>_{\textrm{2}} + \left|a\right>_{\textrm{1}}\left|b\right>_{\textrm{2}}+ \left|b\right>_{\textrm{1}}\left|b\right>_{\textrm{2}},
\end{split}\label{eq:Initial}
\end{equation}
where $\left|a\right>$ and $\left|b\right>$ represent the two optical modes that are mapped onto the two oscillator energy eigenmodes, and $1$ and $2$ label the two photons that are supposed to be distinguishable, \textcolor{RoyalBlue}{by virtue of an extrinsic degree of freedom}. The symbol, $\equiv$ indicates that the subsystems on the two sides of this symbol are different; for example, the entangled subsystems on the left (right) hand side above are modes (particles).

\textcolor{RoyalBlue}{Notice that if photons 1 and 2 were assumed to be identical, then the middle term of Eq. (\ref{eq:Initial}) would have transformed as: $\left|1\right>_{\textrm{a}}\left|1\right>_{\textrm{b}} \equiv \left|a\right>_{\textrm{1}}\left|b\right>_{\textrm{2}} + \left|a\right>_{\textrm{2}}\left|b\right>_{\textrm{1}}$ (up to normalization constants). This appearance of entanglement, as a consequence of the photons' being identical bosons and conforming to exchange symmetry, is a phenomenon known as \emph{entanglement due to symmetrization} \cite{peres2002quantum, dalton2017quantum}. I have described this phenomenon in Fig. \ref{fig:f2}(a), after mapping the spatial, optical modes onto the bound energy eigenmodes of a quantum potential, specifically, a simple harmonic potential. Consequently, one might represent the above state as: $\left|a\right>_{\textrm{1}}\left|b\right>_{\textrm{2}} + \left|a\right>_{\textrm{2}}\left|b\right>_{\textrm{1}} \mapsto \psi\left(x\right)\chi\left(y\right) + \psi\left(y\right)\chi\left(x\right)$, where $\psi\left(x\right)$ and $\chi\left(y\right)$ represent two distinct energy eigenfunctions of the potential, and $x$ and $y$ indicate the spatial coordinates of photons $1$ and $2$, respectively. It is well known that this appearance of entanglement is trivial, and not useful for applications, such as quantum information processing with distinguishable qubits. For instance, two-qubit, entangling gate protocols operating with degrees of freedom—which are used to encode the qubits—require that such degrees of freedom are macroscopically distinguishable, so as to distinguish between the control and the target qubits.}

\textcolor{RoyalBlue}{In contrast, consider the situation, as shown in Fig. \ref{fig:f2}(b), of two distinguishable photons, $1$ and $2$, which are delocalized across a harmonic potential and an anharmonic potential. If we assume that two distinct, complete and orthonormal, sets of spatial, optical modes: $\{\left|a\right>, \left|b\right>, \ldots\}$ and $\{\left|a^{\prime}\right>, \left|b^{\prime}\right>, \ldots\}$ are mapped onto the energy eigenfunctions of the simple harmonic potential and an anharmonic potential, which are, $\{\psi\left(x\right), \chi\left(x\right), \ldots\}$ and $\{\psi^{\prime}\left(x\right), \chi^{\prime}\left(x\right), \ldots\}$, respectively, then such an arrangement can support genuinely entangled states of two photons, for instance, $\left|a\right>_{1}\left|b\right>_{2} + \left|a^{\prime}\right>_{1}\left|b^{\prime}\right>_{2} \mapsto \psi\left(x\right)\chi\left(y\right) + \psi^{\prime}\left(x\right)\chi^{\prime}\left(y\right)$; see, for example, below for a scheme that describes how such a state can be prepared. Crucially—in this example, as well as in the discussion below—I am considering distinguishable photons that occupy distinguishable modes. Such photons are assumed to be distinguishable, due to an additional, external, photonic degree of freedom, for instance, orbital angular momentum, or an extrinsic, translational degree of freedom.}

To be as consistent as possible, in this simple \emph{Gedankenexperiment}, with how optical fields are quantized in the second quantization approach, let us assume that the single-photon states, such as $\left|1\right>_{\textrm{a}}$ and $\left|1\right>_{\textrm{b}}$, map only onto the single-quantum oscillator eigenmodes. The zero- and two-photon states, such as $\left|0\right>_{\textrm{a}}$, $\left|0\right>_{\textrm{b}}$, $\left|2\right>_{\textrm{a}}$, and $\left|2\right>_{\textrm{b}}$, are assumed to map onto distinct oscillator modes, which, presumably, can be adiabatically eliminated from this problem, because of their distinct eigenenergies. Consequently, as depicted in Fig. \ref{fig:f3}(b), let us focus on the middle term, $\widetilde {\left|\Psi\right>_{i}} = \left|1\right>_{\textrm{a}}\left|1\right>_{\textrm{b}} \equiv \left|a\right>_{\textrm{1}}\left|b\right>_{\textrm{2}},$ which corresponds to two photons occupying two distinct modes.

\textcolor{RoyalBlue}{The formulation of the following scheme is partly inspired by the analysis of the Young's double-slit diffraction experiment, with single quanta, in the classic textbook of Schiff \cite{Schiff_1955_Diffraction}. First, however, I wish to provide a key clarification. Typically, detectors and amplifiers are macroscopic, classical objects, and, consequently, quantum mechanical wavefunctions cannot be assigned to such objects, especially, in accordance with the standard, statistical interpretation of quantum mechanics, specifically, the Copenhagen interpretation of quantum mechanics according to Bohr. Nonetheless, in the scheme below, inspired by Schiff's analysis, the detector is taken to be an atom, initially in its ground state, that can be treated as a two-level system. Consequently, the detector, in this particular situation, is a microscopic, measuring apparatus. Also, let us suppose the amplifier to be a linear, phase-invariant—also known as phase-insensitive or phase-preserving—amplifier, whose operation can be fully described by the principles of quantum mechanics. It is reasonable to suppose that a single, macroscopic quantum wavefunction can be assigned to such a quantum optical, phase-coherent amplifier, akin to how a single, macroscopic quantum wavefunction can be assigned to a Bose-Einstein condensate of atoms that is configured to function as a matter wave amplifier, or a matter wave interferometer. Therefore, wavefunctions will be assigned to amplifiers and detectors, in the scheme below. Second, a more rigorous analysis of this scheme, without assigning wavefunctions to amplifiers and detectors, and through the lens of quantum pre-measurement followed by detection, is given in Sec. \ref{sec:Qpm}.}

Let us now deploy a quantum mechanical random process that we will utilize to create macroscopic quantum superpositions. To accomplish this creation, we can introduce an \emph{ancillary} quantum mechanical system, specifically, photon 3, in Fig. \ref{fig:f3}, propagating through the Young's double-slit experimental setup [see Fig. \ref{fig:f3}(a)]. As is well known, the single photon—as it transits through this apparatus—interferes with itself, and, consequently, is described by the superposition: $ \left|\Psi\right>^{(\textrm{photon 3})} \approx \alpha\left|\Psi\right>^{(\textrm{photon 3})}_{\textrm{Top Slit}} + \beta\left|\Psi\right>^{(\textrm{photon 3})}_{\textrm{Bottom Slit}}$ of having gone through the top and the bottom slits.

Assume, furthermore, that the screen is photo-conducting, and is without a lower half—so as to further emphasize that, unlike in the classical case, the position on the screen, for example, the upper half, where the photon is detected, is uncorrelated with the slit of entry, for example, the top slit. \textcolor{RoyalBlue}{Three clarifying comments are in order here, so as to give a more thorough justification for this design choice. First, exactly identical results will be obtained, if only the lower half, as opposed to the upper half, were present, as a consequence of the symmetry of the setup; see, for example, the dotted, horizontal line in Fig. \ref{fig:f3}(a). Second, the lack of correlation between the detection location and the slit originates, fundamentally, from the interference of the single photon with itself—as it transits through the double-slit apparatus—and, therefore, should be independent of the spatial extent of the screen, for instance, if only the upper half were present, or if both halves were utilized. In essence, this lack of correlation stems from the fact that photons behave as quantum mechanical probability amplitudes, as opposed to classical particles. Third, and most interestingly, the above design choice was inspired by a subtle, albeit striking, difference between the predictions of Bohmian mechanics and quantum mechanics, for the outcome of the double-slit experiment. The de Broglie-Bohm pilot wave theory, which is the paradigmatic form of self-consistent, deterministic, hidden variable theories, postulates the existence of hypothetical point corpuscles \footnote{Harvey R. Brown and David Wallace, in \cite{Bohmian2}, makes an excellent argument as to why the particles in the quantum theory of de Broglie and Bohm should be called as hypothetical point corpuscles.} over and above the usual wave function of quantum mechanics. Notice, again, the axis of symmetry in Fig. \ref{fig:f3}(a). No photon, according to such a theory, going through the top slit crosses this axis of symmetry, and the same is true for the photon going through the bottom slit. Consequently, a photon that lands on the top half of the screen must have come through the top slit; likewise, a photon that lands on the bottom half of the screen must have come through the bottom slit \cite{Bohmian1, Dürr2009}. To summarize, the above design choice utilizes not only the usual lack of correlation—as compared with classical mechanics, in which the detection location, for particles, is highly correlated with the slit—but also an enhanced lack of correlation over and above Bohmian mechanics, in which the half-plane of detection is correlated with the slit of entry. I would like to point out that the removal of the lower half of the screen, so as to emphasize, or, possibly, enhance this lack of correlation, is neither an essential nor a crucial feature of this scheme, but is merely an added feature.}

The screen performs a \emph{measurement} of the position of the photon. Now—assuming ideal \emph{detection} and quantum efficiencies—the overall state of the detector and amplifier system, which has access only to photons impinging on the upper half, can be written as the superposition: $ \left|\Psi\right>^{(\textrm{detector and A})} \approx \alpha^{\prime} \left|\Psi\right>^{(\textrm{detector and A})}_{\textrm{photo-e}^{-}} + \beta^{\prime} \left|\Psi\right>^{(\textrm{detector and A})}_{\textrm{No} \; \textrm{photo-e}^{-}}$ of having registered and not registered the single photon as a single photo-electron. \textcolor{RoyalBlue}{Notice that the detection of photon 3, or lack thereof, is described as having created a macroscopic superposition of the detector-amplifier system. Single-photon detection, however, typically collapses the wavefunction. Nevertheless, the formation and the maintenance of the macroscopic quantum superposition can be explained, using the notion of quantum pre-measurement. Actually, the above quantum state is intended to convey a genuine, coherent, linear superposition of the event in which photon 3 is detected, and the one in which photon 3 is not detected; specifically, a state that describes the process of quantum pre-measurement. As is well known, the interaction of a photon with a classical, macroscopic detector involves a fundamentally irreversible and decoherent process, which gives rise to, for example, a classical mixture of an event corresponding to detection and an event corresponding to an absence of detection. Conversely, however, the interaction of a photon with a single, detector atom, which can be treated as a two-level quantum system, gives rise to a coherent linear superposition of an excited atom (and no photon) and an unexcited atom (and a single photon). This state describes the situation preceding the actual measurement upon, or the actual detection of, the photon. Therefore, the words, \emph{measurement} and \emph{detection}, in this particular section, have to be understood in the light of the above clarification.}

Assuming that the regions of photo-detection and collection of photo-electrons are sufficiently macroscopic, the probability amplitude, $\alpha^{\prime} \propto P + iR,$ akin to the expression for the transition probability amplitude of a macroscopically delocalized detector atom for this diffraction experiment \cite{Schiff_1955_Diffraction}. $P + iR$ describes the spectral representation of the electromagnetic field propagating from the source to the detector.

The probabilistically ejected photo-electron is amplified \textcolor{RoyalBlue}{by the phase-coherent amplifier, A,} thereby producing a \textcolor{RoyalBlue}{quantum} photo-current that drives an anharmonic potential controller, which controls the anharmonicity parameter, $\lambda$; for example, no photo-current corresponds to the initially harmonic potential. Consequently, the modal wavefunctions are now described as macroscopic superpositions of the energy eigenfunctions of a harmonic oscillator, and their perturbed counterparts corresponding to an anharmonically altered oscillator.

In essence, the perturbation transforms the state, $\widetilde {\left|\Psi\right>_{i}}$ to this macroscopic superposition:
\begin{equation}
\begin{split}
\widetilde{\left|\Psi\right>_{f}} &= \gamma \left|a\right>_{1}\left|b\right>_{2} + \delta \left|a^{\prime}\right>_{1} \left|b^{\prime}\right>_{2}\\
&\equiv \gamma \left|1\right>_{a}\left|1\right>_{b} + \delta \left|1\right>_{a^{\prime}} \left|1\right>_{b^{\prime}},
\end{split}\label{eq:timescales}
\end{equation}
where, as shown in Fig. \ref{fig:f3}(c), $\left|a^{\prime}\right>$ and $\left|b^{\prime}\right>$ are the corresponding energy eigenmodes of the anharmonic potential. Of note, $\widetilde{\left|\Psi\right>_{f}}$ implies entangling interactions between its subsystems, which are photons. One can imagine this final state as being analogous to: $\gamma^{\prime} \left|\uparrow_{0}\right>_{1}  \left|\downarrow_{0}\right>_{2} +  \delta^{\prime}  \left|\uparrow_{\theta}\right>_{1}  \left|\downarrow_{\theta}\right>_{2}$ in an equivalent spinor representation, where the spin rotation by the angle, $\theta$ models turning on $\lambda$.

\textcolor{RoyalBlue}{Physically, the above state transformation makes use of the well-known notion of amplification of quantum mechanical superpositions. For instance, if a macroscopic superposition of the amplifier-detector system were not formed—such as in the case of a classical mixture of an event corresponding to detection of photon 3, and an event corresponding to an absence of detection of photon 3—then the subsequently collapsed term, $\left|\Psi\right>^{(\textrm{detector and A})}_{\textrm{No} \; \textrm{photo-e}^{-}}$ would have corresponded to the initial term, $\left|a\right>_{1}\left|b\right>_{2}$, whereas the collapsed term, $\left|\Psi\right>^{(\textrm{detector and A})}_{\textrm{photo-e}^{-}}$ would have generated $\left|a^{\prime}\right>_{1} \left|b^{\prime}\right>_{2}$.}

\textcolor{RoyalBlue}{To model the effects of the anharmonic potential controller, one could write down the following two-initial-state, two-particle, effective Hamiltonian:
\begin{equation}
\begin{split}
\hat{H} = &-\hbar\lambda\left( \left|a^{\prime}\right>_{1}\left<a\right|_{1} \otimes \left|b^{\prime}\right>_{2}\left<b\right|_{2}\right)\\
&-\hbar\lambda\left( \left|a\right>_{1}\left<a^{\prime}\right|_{1} \otimes \left|b\right>_{2}\left<b^{\prime}\right|_{2}\right),
\end{split}\label{eq:Hamiltonian}
\end{equation}
where the initially harmonic energy eigenstates, $\left|a\right>$ and $\left|b\right>$ are supposed to smoothly transform into the anharmonically perturbed energy eigenstates, $\left|a^{\prime}\right>$ and $\left|b^{\prime}\right>$, and vice versa; and the particles, 1 and 2 are assumed to make no quantum jumps, either between $\left|a\right>$ and $\left|b\right>$, or between $\left|a^{\prime}\right>$ and $\left|b^{\prime}\right>$. To elucidate how such a Hamiltonian might reliably generate the desired entangled state, consider the following simple, illustrative example. For small values of the perturbation parameter, $\lambda$ and short times, $t$, the initial state, $\widetilde {\left|\Psi\right>_{i}}$ will evolve as:
\begin{equation}
\begin{split}
\hat{U}\widetilde {\left|\Psi\right>_{i}} &= e^{-i\hat{H}t/\hbar}\widetilde {\left|\Psi\right>_{i}}\\
&\approx \left(1-i\hat{H}t/\hbar\right)\widetilde {\left|\Psi\right>_{i}}\\
&=\left|a\right>_{1}\left|b\right>_{2} + i\lambda t \left|a^{\prime}\right>_{1} \left|b^{\prime}\right>_{2}.
\end{split}\label{eq:Unitary}
\end{equation}}

In practice, a probabilistically triggered, time-dependent analogue of the above perturbation should also realize the same effect; however, this perturbation should follow the following hierarchy of timescales, such that the modes, $\left|a\right>$ and $\left|b\right>$ adiabatically evolve into $\left|a^{\prime}\right>$ and $\left|b^{\prime}\right>,$ respectively, and no transitions are induced:
\begin{equation}
\begin{split}
\hbar/\Delta E \ll \hbar/\hat{\widetilde H} \approx \widetilde \omega \ll t_{\textrm{meas}},
\end{split}\label{eq:time}
\end{equation}
where $\Delta E$ is the energy difference between the modes, $\hat{\widetilde H}$ describes the perturbation strength, and $t_{\textrm{meas}}$ is the timescale for carrying out photon number or energy measurements for detecting the inter-particle entanglement of the final state.

Any realization of this \emph{Gedankenexperiment} will have two requisites, namely: (A) an \emph{ancillary} quantum mechanical subsystem has to be coupled to the entangling system; and (B) a quantum measurement has to be made on this subsystem—as are required for entangling photons that have never interacted in the past \cite{PhysRevLett.80.3891}. Hence, I have argued that mode-mode entangled states can be transformed into particle-particle entangled states by anharmonically and controllably—by a quantum mechanical random process—deforming an initially harmonic potential. \textcolor{RoyalBlue}{In particular, in the context of the present scheme, the term, ``particles'' signifies the occupancies of the spatial optical modes. Consequently, one could also interpret this scheme as a quantum measurement-based entangling protocol that transforms the entanglement between two arbitrary, bosonic, optical modes into entanglement between their constituent, characteristic, spatial modes.}

\textcolor{RoyalBlue}{Notably, this final photon entanglement differs from the path entanglement that is generated, using traditional methods through mode interference, since the paths of these photons—corresponding to, for example, distinct interferometric paths within a multi-mode interferometer—can be described and modeled as genuine spatial optical modes. I would also like to emphasize that the physical degrees of freedom—that encapsulate the entanglement following the conversion process—are the photon occupation numbers within a conserved total photon occupation number subspace. I speculate, however, that it is certainly possible to propose and implement similar schemes and protocols that convert the wavefunctional entanglement into other physical degrees of freedom that are intrinsic to the photon, and that are fully specified by the wavefunctions, or, more specifically, the state vectors of the photons, together with the appropriate Hermitian operator for the observable. Proposing such schemes and protocols, however, is beyond the scope of this paper.}

\textcolor{RoyalBlue}{The current scheme is based on two strong assumptions, both of whose experimental feasibilities, in the context of current technological capabilities, impose severe limitations on the practical utility of this scheme. First, as has been discussed earlier, the photons are assumed to be distinguishable due to an extrinsic degree of freedom, for instance, orbital angular momentum, or some other rotational state. In particular, experiments have already shown the possibility of accessing and manipulating the orbital angular momenta of optical photons \cite{OAMZeilinger2001, 10.1098/rsta.2015.0442}. Second, the efficiency of detecting the ancillary photon 3 is assumed to be unity, and the formation of the macroscopic superposition of the detector-amplifier system is assumed to be ideal, without any accompanying errors. Section \ref{sec:Photodetect} discusses how this issue can be addressed for non-ideal photodetection efficiencies in the context of realistic, present-day detectors.}

\section{A Comparison with Entanglement Swapping}\label{sec:Swap}

The above-mentioned entangling scheme has several striking similarities to the well-known procedure of entanglement swapping \cite{PhysRevLett.70.1895, PhysRevLett.71.4287, PhysRevA.57.822, PhysRevLett.80.3891}, such as projecting the state of two particles onto an entangled state, by making a measurement on an ancillary system. In fact, the fundamental principle of operation underlying both approaches is essentially identical, namely that entanglement is a physical resource that can be transferred between two distinct, non-interacting, uncoupled systems. In what follows, I will briefly review the main idea of entanglement swapping, which I will then compare and contrast with my proposal.

Consider two quantum subsystems, each comprising a pair of initially entangled photons; assume the absence of any direct, physical interactions, or dynamical couplings whatsoever between these subsystems. Now, a suitable kind of projective measurement, such as a joint Bell-state measurement, on two of such photons, each drawn from a distinct initially entangled pair, will project the other two into an entangled state. Moreover, the generation of the results of the Bell-state measurement on two particles heralds that the other two particles have been entangled, thereby allowing, for example, the performance of \emph{event-ready detections} of the entangled particles \cite{PhysRevLett.80.3891}.

Typically, various generalizations of the entanglement swapping scheme entail using pairs of initially entangled photons and projecting the state of two initially unentangled particles onto an entangled state. In contrast, the present scheme transforms the entanglement between optical modes (the wavefunctions) into entanglement between the physical properties of the photons (the wavefunctional degrees of freedom), which are described by these modes. As was pointed out in Ref. \cite{PhysRevLett.80.3891}: \emph{One could have many different kinds of entanglements to begin with, perform various different measurements, and obtain various kinds of entanglement for the
emerging particles.} The present approach, therefore, is a form of entanglement swapping that transforms mode-mode entanglement into particle-particle entanglement.

\textcolor{RoyalBlue}{Nonetheless, unlike the above approach, standard entanglement swapping protocols rely on projective measurements without external potentials. The introduction of a controlled anharmonic potential signifies a substantial deviation from this standard framework of entanglement swapping. Conceptually, as I have emphasized above, the key difference between this approach and the usual entanglement swapping protocols is that while the former transforms and projects \emph{mode-mode entanglement} onto \emph{particle-particle entanglement}, typically, the latter transforms and projects \emph{particle-particle entanglement} onto a similar form of entanglement, namely, \emph{particle-particle entanglement}. Operationally, a controlled anharmonic potential appears to be essential, so as to adiabatically manipulate the wavefunctions of the mode-mode entangled states, and reliably and smoothly transform such states into genuinely entangled states of particles of light. Finally, I wish to point out that while my scheme supplements projective measurement on an ancillary photon with a controlled anharmonic potential, it could certainly be possible to supplement projective measurements with other kinds of quantum circuits and systems that reliably and smoothly alter the modal wavefunctions, so as to generate identical final entangled states. The exploration of such additional schemes and protocols, however, is beyond the scope of this paper.}

\section{Consequences of Non-Ideal Photodetection Efficiencies}\label{sec:Photodetect}

The primary purpose of the simple, ideal \emph{Gedankenexperiment}, described in Sec. \ref{sec:Equivalence}, is to point out the possibility of transforming entangled states of modes into entangled states of particles. From a practical point of view, however, the main disadvantage of any realistic scheme, based on this approach, would be its inescapable reliance on successful photodetection of the ancillary photon. Photodetection, typically, has an efficiency, $\eta < 1 $, and, therefore, it is important to consider whether such a scheme would survive realistic photodetection conditions.

An inability to register the ancillary photon that lands on the top half of the screen B (see Fig. \ref{fig:f3}), due to non-ideal photodetection conditions, can be modeled as an error in forming the state, $ \left|\Psi\right>^{(\textrm{detector and A})} \approx \alpha^{\prime} \left|\Psi\right>^{(\textrm{detector and A})}_{\textrm{photo-e}^{-}} + \beta^{\prime} \left|\Psi\right>^{(\textrm{detector and A})}_{\textrm{No} \; \textrm{photo-e}^{-}}$; such single-photon-loss errors will cause errors in the probability amplitudes, $\alpha^{\prime}$ and $\beta^{\prime}$, and, consequently, in $\gamma$ and $\delta$. Therefore, the correct output state, $\widetilde{\left|\Psi\right>_{f}}$ will not be formed, and the maximum efficiency of conversion from $\left|\Psi\right>_{i}$ to $\widetilde{\left|\Psi\right>_{f}}$ will be limited by $\eta$.

Realistically, the detection screen B can be implemented by a single-photon-sensitive EMCCD (Electron Multiplying Charge-Coupled Device) camera. State-of-the-art EMCCD cameras, typically, have quantum efficiencies (equivalent to the photodetection efficiency, above), $\eta \approx 0.9$ that can range up to $\approx 0.95$ in the visible wavelengths. Moreover, these cameras have advanced features, such as vacuum thermoelectric cooling and electronic optimization of clock-induced charge, for minimizing spurious photodetection events and the dark noise.

One intriguing idea might be to borrow ideas and concepts from standard quantum error-correction theory, so as to detect and mitigate these ancillary-photon-loss errors. For example, a key finding from recent experiments with neutral atom-based quantum processors is that atom-loss-type leakage errors, if detectable and convertible to erasure errors, are easier to correct than other generic classes of unknown errors \cite{PhysRevX.12.021049, Wu_Erasure, PRXQuantum.5.020355, PRXQuantum.5.040343}. Inspired by these ideas pointing to the feasibility of correcting single-quantum-losses, I wish to propose a quantum error-correcting-based strategy, for mitigating the above-described ancillary-photon-loss errors, that is based on classical control, and that is specific to the experiment in Sec. \ref{sec:Equivalence}.

Let us imagine that the single-photon source, S is deterministically triggered by a clock signal. The frequency of the clock signal, and, therefore, the single-photon emission rate is kept sufficiently low, such that the time interval between the emissions of two successive photons is greater than the sum of the time it takes the single photon to travel from S to the screen, the time for this photon to be detected and registered, and the time required for a classical, logical \texttt{AND} operation that is described in the next paragraph. Unlike the experiment described in Sec. \ref{sec:Equivalence}, let us not discard the bottom half of the screen, but rather implement the entire screen with a EMCCD camera that is symmetrically positioned with respect to the dotted horizontal line shown in Fig. \ref{fig:f5}(a). Let us now equip this camera with two measurement read-out channels, R1 and R2, such that R1 outputs the photocurrent generated due to single-photon-detection events in the top half of the camera screen, and R2 outputs the photocurrent generated due to single-photon-detection events in any part of the camera screen; such a situation can be readily arranged by writing a suitable piece of code with the software that controls the camera. \textcolor{RoyalBlue}{Simply put, for every single-photon-detection event, the software will, based on the information from the two measurement read-out channels, R1 and R2, will record two classical bits of information: the first signifying that a successful detection event has occurred; and the second indicating whether this detection has occurred in the top half of the screen.} The channel, R1 is connected to the amplifier, A and the anharmonic potential controller, thereby providing the probabilistic triggering mechanism required in Fig. \ref{fig:f3}(b).

In contrast, the channel, R2 and the clock signal are inputted to a classical, logical \texttt{AND} circuit, and the output is monitored for a window of time equal to the period of the clock signal. If the output is $1$, then no action is taken; however, if the the output is $0$, then it is assumed that the ancillary photon has been lost, and the entire cycle of quantum operation, which commences with the emission of the ancillary photon and concludes with the formation of $\widetilde{\left|\Psi\right>_{f}}$, is aborted. A new cycle of quantum operation on a freshly-prepared, input, entangled state of modes starts with the next clock cycle.

Importantly, the above classical operation should be carried out only after the formation of the macroscopic superposition of the amplifier and the detector, so as not to interfere with the probabilistic altering of the anharmonicity of the potential. In essence, while the efficiency of conversion from $\left|\Psi\right>_{i}$ to $\widetilde{\left|\Psi\right>_{f}}$ remains unaltered, the output entangled state of photons becomes significantly error-free, at the cost of reduced speed of operations, which is, fundamentally, determined by the single-photon-emission rate.

\textcolor{RoyalBlue}{It is important to clarify the practical costs, which are associated with the repeated restarts that are necessitated by the above-described logical circuit abort strategy. With regard to resource overhead, this abort strategy requires that the number of times the initial state, $\left|\Psi\right>_{i}$ is prepared, is increased by a factor of $1/\eta$, as the rate of abortion is $\eta \times 100 \%$. Additionally, this strategy requires adding two equipment—such as a clock signal generator and the classical, logical, \texttt{AND} gate—as well as reconfiguring the EMCCD camera, so as to have two measurement read-out channels. Finally, I will conclude this section by offering a few comments about the increased time complexity. In the absence of the abort strategy, one might argue that the total operation time is $\sim \mathcal{O}\left(t_{\textrm{r}} + t_{\textrm{A}} + t_{\lambda}\right)$, where $t_{\textrm{r}}$ is the response time of the detector, and $t_{\textrm{A}}$ and $t_{\lambda}$ are the time scales of operation of the amplifier and the anharmonic potential controller, respectively. This total operation time increases, as a consequence of introducing the abort strategy, to $\sim \mathcal{O}\left(t_{\textrm{r}} + t_{\textrm{A}} + t_{\lambda} + t_{\texttt{AND}}\right)$, where $t_{\texttt{AND}}$ is the timescale of operation of the classical, logical, $\texttt{AND}$ gate. More accurately, this increased operation time can be modeled as $\sim \mathcal{O}\left(t_{\textrm{clock}}\right)$—where $t_{\textrm{clock}}$ is the time period of the clock signal, and $t_{\textrm{clock}} > t_{\textrm{r}} + t_{\textrm{A}} + t_{\lambda} + t_{\texttt{AND}}$—since the timescale of formation of the final, error-corrected, entangled state of photons is fundamentally limited by $t_{\textrm{clock}}$. As mentioned earlier, the clock rate is identical to the single-photon-emission rate.}

\begin{figure*}[ht]
\includegraphics[width=\textwidth]{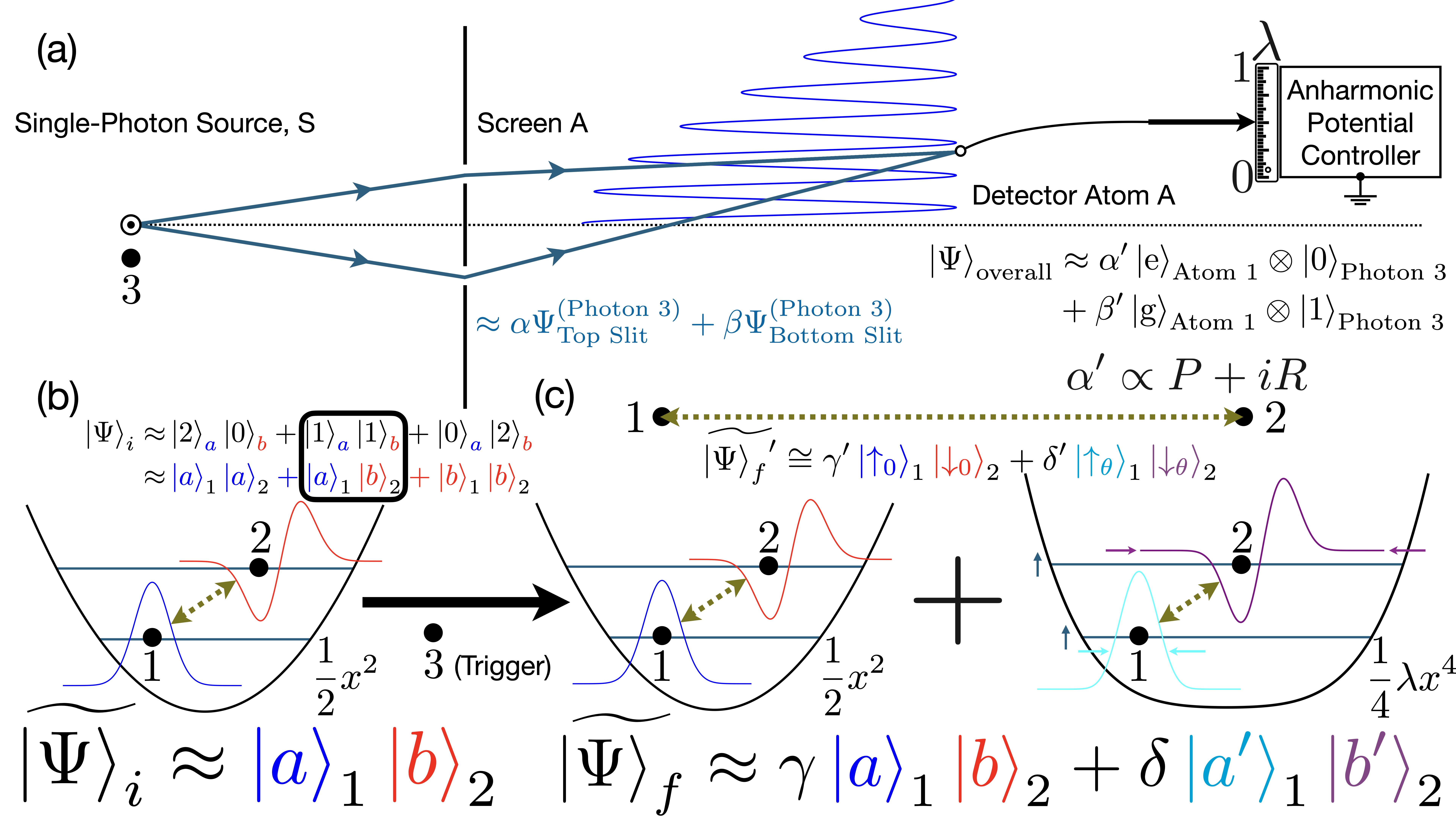}
\caption{\label{fig:f4} \textcolor{RoyalBlue}{\textbf{A form of the \emph{Gedankenexperiment} shown in Fig. \ref{fig:f3} that has been further simplified, so as to clarify and highlight the importance of the notion of quantum pre-measurement in analyzing the entangling scheme described in Sec. \ref{sec:Equivalence}.} (a) A version of the well-known Young's double-slit diffraction experiment, in which the superposition of an ancillary photon (labeled as 3)—of having traveled through the top and the bottom slits—is transformed into a superposition of product states of the macroscopically-delocalized, two-level atom, A and the single photon, 3. Notice that this atom acts as a quantum meter, namely, a microscopic, measuring apparatus, which is treated quantum mechanically, and the overall excitation number of this atom is the relevant pointer observable. The ground and the excited states of this two-level atom are supposed to correspond to an unionized and an ionized atomic state, respectively. Subsequent to the formation of such a superposition of an excited atom—with an accompanying ejected photo-electron—and a non-excited atom—without any ejected photo-electron—a quantum mechanical photocurrent is produced that probabilistically triggers the anharmonic potential controller. The parameter, $\lambda$ indicates the degree of triggering. The curved black curve that connects the detector atom and the anharmonic potential controller, along with the grounding symbol on the latter, is supposed to represent a quantum mechanical photo-current, specifically, the flow of a quantum mechanical state of photo-electrons through a closed circuit. (b) and (c) Same as Figs. \ref{fig:f3}(b) and (c), respectively. These sub-figures have been reproduced here, so that the reader can visualize the entire entangling scheme in one figure.}}
\end{figure*}

\section{Quantum Pre-measurement followed neither by Decoherence nor by Detection}\label{sec:Qpm}

\textcolor{RoyalBlue}{The purpose of this section is to reconsider and re-analyze the \emph{Gedankenexperiment}, as described in Sec. \ref{sec:Equivalence}, using the notion of quantum pre-measurement. A key issue that one inadvertently encounters, while analyzing the said \emph{Gedankenexperiment}, is that one has to, inexorably, assign quantum mechanical wavefunctions to detectors and amplifiers, which are, usually, classical and macroscopic objects. In this section, I will further simply this \emph{Gedankenexperiment} to its simplest and most essential elements. The screen and the conventional detector is replaced by a single detector atom, which is treated quantum mechanically and as a two-level quantum system; furthermore, there is no longer any amplifier. In what follows, I will briefly survey the usual theory of quantum pre-measurement followed by decoherence—and/or detection—and subsequently justify the entangling scheme of Sec. \ref{sec:Equivalence}, by utilizing the concept of a quantum pre-measurement of a quantum system by a microscopic, well-isolated, quantum meter.}

\textcolor{RoyalBlue}{Typically, in any theoretical formulation of the process of quantum measurement \cite{PhysRevD.26.1862, RevModPhys.75.715, PhysRevLett.126.130402}, the Hilbert space of the entire universe is factorized into those of the quantum system, S; the measuring apparatus, A; and the rest of the universe, or the environment, E. The quantum system, S is assumed to be small enough, and/or isolated enough, such that its initial state, or the so-called ready state, can be described as a genuine, coherent, linear superposition of two or more eigenstates of the relevant quantum observable. Usually, A is assumed to have multiple macroscopic, accessible degrees of freedom, whereas E is supposed to contain multiple microscopic, hidden or distant degrees of freedom—corresponding to the rest of the environment—that we are not keeping track of.}

\textcolor{RoyalBlue}{First, during a process known as a pre-measurement, A, as a consequence of unitary time evolution, measures S and becomes entangled with S. The value of the pointer observable of A becomes correlated with the corresponding eigenvalue of the observable of S, which A is supposed to measure. Notice that A interacts directly with S. Second, if A is sufficiently macroscopic and/or not sufficiently well-isolated, then it gets entangled with the rest of the environment, in a process known as decoherence or environmental entanglement. Physically, the environment is continuously monitoring the combined system of $\textrm{S} + \textrm{A}$, and, consequently, A splits into preferred pointer basis states that are robust with respect to continual monitoring by the environment. In equations, this two-step process can be summarized as:
\begin{equation}
\begin{split}
\left|\Psi\right>_{\textrm{overall}} &= \left(\left|\uparrow\right>_{\textrm{S}} + \left|\downarrow\right>_{\textrm{S}}\right)\left|\textrm{A}_{0}\right>_{\textrm{A}}\left|\textrm{e}_{0}\right>_{\textrm{E}}\\
&\rightarrow \left(\left|\uparrow\right>_{\textrm{S}}\left|\textrm{A}_{\uparrow}\right>_{\textrm{A}} + \left|\downarrow\right>_{\textrm{S}}\left|\textrm{A}_{\downarrow}\right>_{\textrm{A}}\right)\left|\textrm{e}_{0}\right>_{\textrm{E}}\\
&\rightarrow \left|\uparrow\right>_{\textrm{S}}\left|\textrm{A}_{\uparrow}\right>_{\textrm{A}}\left|\textrm{e}_{\uparrow}\right>_{\textrm{E}} + \left|\downarrow\right>_{\textrm{S}}\left|\textrm{A}_{\downarrow}\right>_{\textrm{A}}\left|\textrm{e}_{\downarrow}\right>_{\textrm{E}},
\end{split}\label{eq:Measurement}
\end{equation}
wherein the environmental states become orthogonal: $\braket{\textrm{e}_{\uparrow}|\textrm{e}_{\downarrow}} \approx 0$, within a very short timescale, namely, the decoherence timescale, such that the two pieces of the final, decoherent, overall wavefunction cannot interfere with each other. The process of decoherence is, therefore, sometimes described as having branched the overall wavefunction into causally distinct, non-interacting worlds that are decoherent with respect to each other.}

\textcolor{RoyalBlue}{Now, with the above discussion in mind, let us consider the simplified \emph{Gedankenexperiment}, as shown in Fig. \ref{fig:f4}. The atom, A is assumed to be macroscopically delocalized—that is, not in a definite eigenstate of position—along the diffraction plane, and to behave as a \emph{quantum meter}, namely, a microscopic, measuring apparatus, which is treated quantum mechanically. Moreover, consider this atom to be well-modeled by a two-level system, such that its ground state corresponds to an unionized atomic state and its excited state corresponds to an ionized atomic state, and assume the difference between the excited and the ground energy eigenstates of this atom to be identical to the energy of photon 3. Notice that there are no longer any amplifiers or macroscopic detectors.}

\textcolor{RoyalBlue}{As has been discussed in Sec. \ref{sec:Equivalence}, the ancillary photon 3 can be described by the state: $\approx \alpha \Psi^{\left(\textrm{Photon 3}\right)}_{\textrm{Top Slit}} + \beta\Psi^{\left(\textrm{Photon 3}\right)}_{\textrm{Bottom Slit}}$, as it transits through the double-slit apparatus. In accordance with the analysis of this double-slit diffraction experiment, in the textbook of Schiff \cite{Schiff_1955_Diffraction}, the overall state of the macroscopically delocalized detector atom and the single photon can be described as:
\begin{equation}
\begin{split}
\left|\Psi\right>_{\textrm{overall}} &\approx \alpha^{\prime}\left|\textrm{e}\right>_{\textrm{Atom 1}}\otimes\left|0\right>_{\textrm{Photon 3}}\\
&+ \beta^{\prime}\left|\textrm{g}\right>_{\textrm{Atom 1}}\otimes\left|1\right>_{\textrm{Photon 3}},\\
\end{split}\label{eq:DetectorAtom}
\end{equation}
where $\left|\textrm{e}\right>_{\textrm{Atom 1}}$ and $\left|\textrm{g}\right>_{\textrm{Atom 1}}$ are the two energy eigenstates of the detector atom, and the expression for the probability amplitude, $\alpha^{\prime}$—specifically, the atomic transition probability amplitude—is given in Ref. \cite{Schiff_1955_Diffraction}. Notice that the detector atom acts as a quantum meter and that the total excitation number, of this atom, is the salient pointer observable. Subsequent to the formation of the above superposition of an excited atom—with an accompanying ejected photo-electron—and a non-excited atom—without any ejected photo-electron—a quantum mechanical photocurrent is produced that corresponds to a superposition of the form: $\approx \alpha^{\prime\prime}\Psi_{\textrm{photo-e}^{-}} + \beta^{\prime\prime}\Psi_{\textrm{No photo-e}^{-}}$. This quantum mechanical photocurrent probabilistically triggers the anharmonic potential controller; a value of $\lambda > 0$ is supposed to indicate triggering. Crucially, I am assuming that the anharmonic potential controller is spatially located sufficiently close to the detector atom, such that the anharmonic potential controller is within the region of successful capture of the photo-electron; furthermore, it is assumed that the anharmonic potential controller is suitably grounded, so that a closed circuit is established for the flow of photo-electrons.}

\textcolor{RoyalBlue}{It is essential to emphasize that both the detector atom and the single-electron-based triggering mechanism—that can be realized, for instance, using single-electron turnstile devices \cite{PhysRevLett.64.2691, PhysRevLett.116.166801}— of the anharmonic potential controller are all assumed to be microscopic, quantum mechanical, and sufficiently well-isolated from the rest of the environment, so that there is neither any decoherence nor any detection nor any collapse of the above superpositions. From this point onward, the remainder of the analysis of how the final entangled state of photons is formed is identical to what has been described in Sec. \ref{sec:Equivalence}, specifically, from the presentation of the Hamiltonian in Eq. (\ref{eq:Hamiltonian}) onward.}

\section{Conclusions and Outlook}\label{sec:Conclude}

Broadly construed, photonic entangled states are described as either mode-mode entangled, or photon-photon entangled, where the relevant quantum subsystems are modes (wave-like) and photons (particle-like), respectively. This classification criterion is also applicable to quantum systems of cold, bosonic atoms, molecules, and ions. Mode-mode entangled states are relatively easier to prepare—for example, by amplitude and beam splitting—than photon-photon entangled states; however, the latter is a key prerequisite for optical quantum information processing. In fact, experimentally constructing photon-photon entangled states is one of the grand challenges of quantum optics.

In this paper, I have examined the equivalence of entanglement between optical modes and entanglement between optical photons that are described by such modes, and have concluded that it is, indeed, possible to transform the former into the latter; usually, the latter is harder to realize and is more useful than the former. Specifically, I have shown that systems having four or more spatial modes, and two or more input unentangled photons—such as non-Abelian, quantum holonomic systems \cite{neef2023three, Neef:25, neef2025pairingparticlesholonomies} in which light field modes are in photon number entangled relationships with each other \cite{PhysRevLett.134.080201}—can be used to nonlocally entangle photons. This theoretical demonstration might have implications for crafting protocols and devising algorithms—that, for instance, start with mode-mode entangled states, which are, subsequently, converted to particle-particle entangled states, during a key, intermediate step—with continuous wave, quantum optical systems.

Highly entangled optical systems can be used to construct modular, programmable platforms that are more readily interfaced and integrated with existent long-distance quantum communication networks, which already employ optical photons encoded as flying qubits. Additionally, a future avenue of fundamental research might entail placing the results of this paper on a firmer mathematical grounding, and extending them to quantum systems found in other branches of physics.

\begin{acknowledgments}
I would like to thank Sara Sloman, J. Van Butcher, Chandra Raman, and Xueda Wen, for many delightful and helpful discussions—on related topics—that inspired me to work on this problem. Additionally, I wish to thank Rajesh Gopakumar and Abhishek Dhar, for hosting me at the ICTS--TIFR, Bengaluru, where a part of this paper was written.    
\end{acknowledgments}

\section{Data Availability}\label{sec:Data}

This paper reports a theoretical work, and consequently, no experimental research data were created or analyzed, during the course of this research. In particular, the author confirms that all the relevant information—namely, theoretical calculations, derivations, and accompanying results—that support the findings of this work are available within the article.

\appendix
\section{Entanglement Measurement Contexts in Real and Hilbert Spaces.}\label{sec:Context}

\begin{figure*}[ht]
\includegraphics[width=\textwidth]{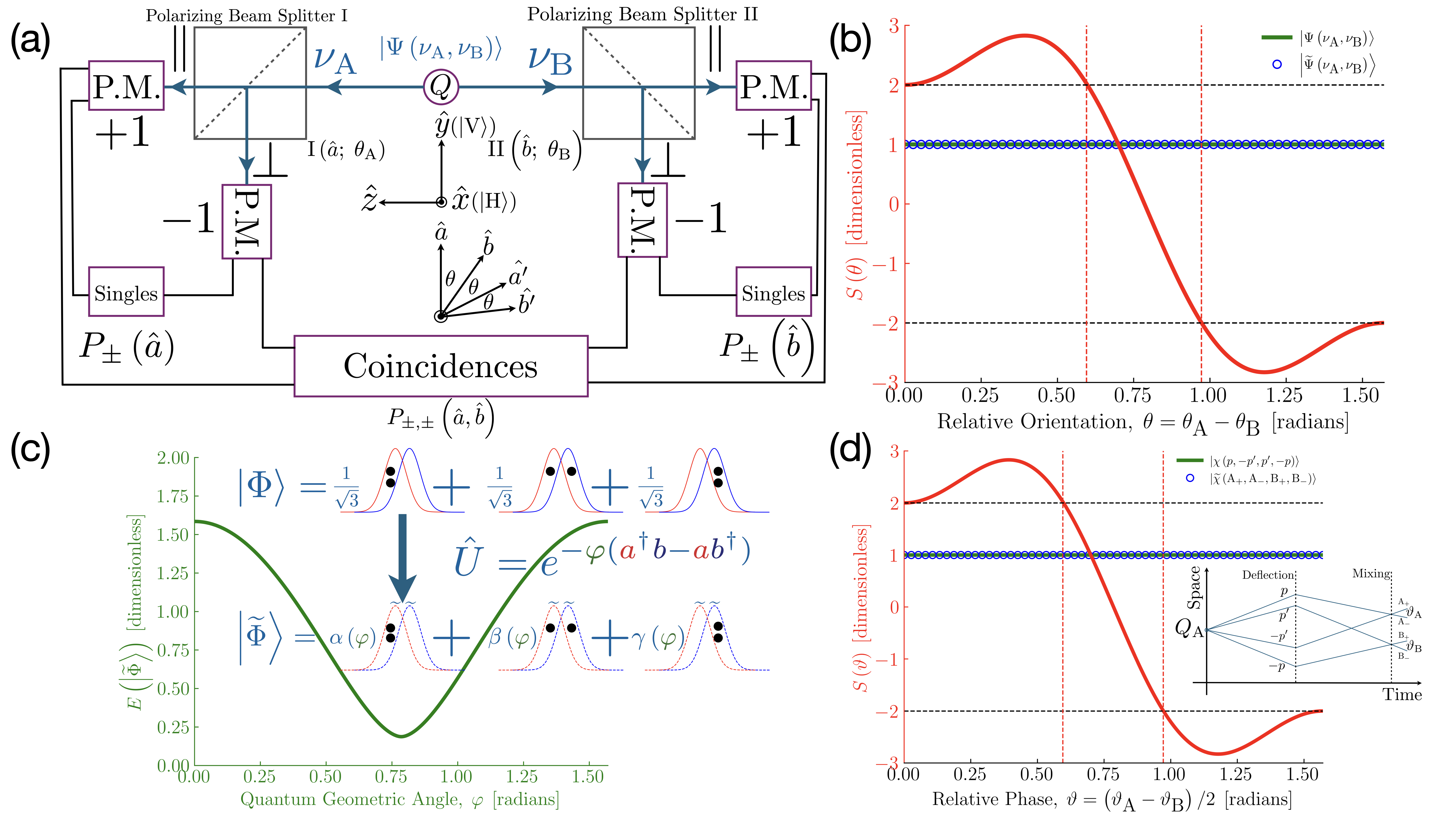}
\caption{\label{fig:f5} \textbf{Effects of relative phases in real and Hilbert spaces on inter-particle and inter-modal entanglement.} (a) An optical re-interpretation of the E.P.R.B. \emph{Gedankenexperiment}, which was originally devised and realized by Aspect \emph{et al.}, and enables the direct measurements of the single-particle, and the joint, two-particle detection probabilities, so as to verify the Bell nonlocality of the two-particle entangled state, $\left|\Psi\left(\nu_{\textrm{A}}, \nu_{\textrm{B}}\right)\right>$. The subsystems are the photons and the entangled property is the angle of linear polarization. The angles, $\theta_{\textrm{A}}$ and $\theta_{\textrm{B}}$ are in the $x-y$ plane. (b) The algebraic sum of the correlation coefficients, $S$ as a function of the relative orientation of the two linear analyzers in real space, $\theta = \theta_{\textrm{A}}-\theta_{\textrm{B}}$ (solid red curve). The solid green line and the hollow blue circles indicate the von Neumann entropies of entanglement of $\left|\Psi\left(\nu_{\textrm{A}}, \nu_{\textrm{B}}\right)\right>$, and its unitarily transformed version, $\left|\widetilde{\Psi}\left(\nu_{\textrm{A}}, \nu_{\textrm{B}}\right)\right>$, respectively. All local, realist, hidden-variables theories, formulated according to the Einsteinian worldview, predict $-2\leq S \leq2$. The violation of the Bell's inequalities on the left and right of the vertical, dashed lines is the telltale signature of quantum nonlocality. (c) The von Neumann entropy of entanglement of a two-particle, two-mode entangled state—where the quantum subsystems are the modes, and the entangled property is the photon occupation number—as a function of the angle of rotation in the Hilbert space, $\varphi$. Notice that a change in the relative angular orientation in the Hilbert space—as opposed to in the real space—can modulate the entanglement entropy. (d). Same as (b), but for a two-particle, four-mode entangled state, where the quantum subsystems are the modes, and the entangled property is the linear momentum of the particle. The inset shows a space-time arrangement that allows the output modes to be interfered two-by-two—with individually controlled phases, $\vartheta_{\textrm{A}}$ and $\vartheta_{\textrm{B}}$ at two distinct spatial locations—so as to verify this state's Bell nonlocality.}
\end{figure*}

To distinguish between the contextual natures of photon-photon and mode-mode entanglement, consider a quantum optical re-interpretation of the E.P.R.B. \emph{Gedankenexperiment}, as shown in Fig. \ref{fig:f5}(a). If we regard the horizontally and vertically polarized, single-photon states, $\left|\textrm{H}\right>$ and $\left|\textrm{V}\right>$ as the basis states—for example, $\left|\textrm{H}\right> = \begin{bmatrix}
1 \\
0 
\end{bmatrix}$ and $\left|\textrm{V}\right> = \begin{bmatrix}
0 \\
1 
\end{bmatrix}$—then the corresponding, unitarily transformed states at the measurement station A (orientation angle, $\theta_{\textrm{A}}$; axis of linear analyzer, $\hat{a}$) are:
\begin{equation}
\begin{split}
\left|+\right>_{\textrm{A}} &= \cos\theta_{\textrm{A}}\left|\textrm{H}\right>_{\textrm{A}}+\sin\theta_{\textrm{A}}\left|\textrm{V}\right>_{\textrm{A}},\\
\left|-\right>_{\textrm{A}} &= -\sin\theta_{\textrm{A}}\left|\textrm{H}\right>_{\textrm{A}}+\cos\theta_{\textrm{A}}\left|\textrm{V}\right>_{\textrm{A}},
\end{split}\label{eq:Transformation_A}
\end{equation}
and similarly, the unitarily transformed states at the measurement station B (orientation angle, $\theta_{\textrm{B}}$; axis of linear analyzer, $\hat{b}$) are:
\begin{equation}
\begin{split}
\left|+\right>_{\textrm{B}} &= \cos\theta_{\textrm{B}}\left|\textrm{H}\right>_{\textrm{B}}+\sin\theta_{\textrm{B}}\left|\textrm{V}\right>_{\textrm{B}},\\
\left|-\right>_{\textrm{B}} &= -\sin\theta_{\textrm{B}}\left|\textrm{H}\right>_{\textrm{B}}+\cos\theta_{\textrm{B}}\left|\textrm{V}\right>_{\textrm{B}}.
\end{split}\label{eq:Transformation_B}
\end{equation}

One can use the above expressions to derive Eq. (\ref{eq:Transformed_Entangled_State}) from Eq. (\ref{eq:Entangled_EPR_State}), and verify that: 
\begin{equation}
\begin{split}
P_{\pm}\left(\hat{a}\right) &= \left|\braket{\widetilde{\Psi}\left(\nu_{\textrm{A}}, \nu_{\textrm{B}}\right)|\pm}_{\textrm{A}}\right|^2 = 1/2,\\
P_{\pm}\left(\hat{b}\right) &= \left|\braket{\widetilde{\Psi}\left(\nu_{\textrm{A}}, \nu_{\textrm{B}}\right)|\pm}_{\textrm{B}}\right|^2 = 1/2,\\
P_{\pm \pm}\left(\hat{a},\hat{b}\right) &= \left|\braket{\widetilde{\Psi}\left(\nu_{\textrm{A}}, \nu_{\textrm{B}}\right)|\pm,\pm}_{\textrm{A,B}}\right|^2 = \frac{1}{2}\cos^2 \theta,\\
P_{\pm \mp}\left(\hat{a},\hat{b}\right) &= \left|\braket{\widetilde{\Psi}\left(\nu_{\textrm{A}}, \nu_{\textrm{B}}\right)|\pm,\mp}_{\textrm{A,B}}\right|^2 = \frac{1}{2}\sin^2 \theta,\\
E\left(\hat{a},\hat{b}\right) &= P_{++}\left(\hat{a},\hat{b}\right) + P_{--}\left(\hat{a},\hat{b}\right)\\ 
&- P_{+-}\left(\hat{a},\hat{b}\right) - P_{-+}\left(\hat{a},\hat{b}\right)\\
&= \cos 2\theta,\\
\end{split}\label{eq:Probabilities}
\end{equation}
where $P_{\pm}\left(\hat{a}\right)$ and $P_{\pm}\left(\hat{b}\right)$ are the single-particle, detection probabilities; $P_{\pm\pm}\left(\hat{a},\hat{b}\right)$ are the two-particle, joint, detection probabilities; $E\left(\hat{a},\hat{b}\right)$ is the coefficient of correlation; and $\theta = \theta_{\textrm{A}}-\theta_{\textrm{B}}$.

Let us now consider two axes of measurement at the measurement station A, $\hat{a}$ and $\hat{a^\prime}$, and two axes of measurement at the measurement station B, $\hat{b}$ and $\hat{b^\prime}$, such that:
\begin{equation}
\begin{split}
\angle \left(\hat{a}, \hat{b}\right) = \angle \left(\hat{b}, \hat{a^{\prime}}\right) &= \angle \left(\hat{a^{\prime}}, \hat{b}\right) = \angle \left(\hat{a^{\prime}}, \hat{b^{\prime}}\right) = \theta,\\
\end{split}\label{eq:Angles1}
\end{equation}
and, as can be seen from Fig. \ref{fig:f5}(a):
\begin{equation}
\begin{split}
\angle \left(\hat{a}, \hat{b^{\prime}}\right) = 3\theta.\\
\end{split}\label{eq:Angles2}
\end{equation}
Therefore, the well-known algebraic sum of four correlation coefficients, $S$—involving four measurements in four distinct orientations—can be written as: 
\begin{equation}
\begin{split}
S &= E\left(\hat{a},\hat{b}\right)-E\left(\hat{a},\hat{b^{\prime}}\right)+E\left(\hat{a^{\prime}},\hat{b}\right)+E\left(\hat{a^{\prime}},\hat{b^{\prime}}\right)\\
&= 3\cos 2\theta - \cos 6\theta.
\end{split}\label{eq:AlgebraicSum}
\end{equation}
Figure \ref{fig:f5}(b) graphically summarizes all these results. Additionally, the entropies of entanglement are found to have no dependence on the relative orientations of the analyzers in space. In contrast, as shown in Fig. \ref{fig:f5}(c), a rotation in Hilbert space can modulate the expansion coefficients, and, therefore, the degree of entanglement of a mode-mode entangled state. Recently, an approach based on non-Abelian holonomy has been devised to access and tune the angle of this rotation, $\varphi$ in a deterministic fashion in a real-world, laboratory setting \cite{PhysRevLett.134.080201}. Identical results are obtained in Figs. \ref{fig:f5}(b) and \ref{fig:f5}(c), if the R{\'e}nyi entropies are computed, instead of the von Neumann entropies.

Finally, consider the measurement context, as realized by the two-atom, four-momentum-mode interferometer \cite{PhysRevLett.119.173202} shown in the inset of Fig. \ref{fig:f5}(d), for detecting the entanglement between the modes of neutral atoms. The four modes are made to interfere two-by-two at two distinct spatial locations, such that the joint detection probabilities are directly accessible and measurable. Specifically, the deflection and mixing are achieved by Bragg diffraction.

Effectively, this interferometer transforms the input state \cite{PhysRevLett.119.173202}:
\begin{equation}
\begin{split}
\left|\chi\left(p, -p^{\prime}, p^{\prime}, -p\right)\right> = \frac{1}{\sqrt{2}}\Bigl\{\left|p, -p\right>_{1,2}+\left|p^{\prime}, -p^{\prime}\right>_{1,2}\Bigr\},
\end{split}\label{eq:Input_Interfere_State}
\end{equation}
where $p$ and $p^{\prime}$ are the two atomic linear momenta, and $1$ and $2$ label the two atoms moving in opposite directions, to the output state \cite{PhysRevLett.119.173202}:
\begin{equation}
\begin{split}
\left|\widetilde{\chi}\right> &= \frac{1}{2\sqrt{2}}\Bigl\{-ie^{i\vartheta_{\textrm{B}}}\left(e^{i\left(\vartheta_{\textrm{A}}-\vartheta_{\textrm{B}}\right)}+1\right)\left|A_{+}, B_{+}\right>\\
&+ \left(e^{i\left(\vartheta_{\textrm{A}}-\vartheta_{\textrm{B}}\right)}-1\right)\left|A_{+}, B_{-}\right>\\
&+ \left(e^{-i\left(\vartheta_{\textrm{A}}-\vartheta_{\textrm{B}}\right)}-1\right)\left|A_{-}, B_{+}\right>\\
&-
ie^{-i\vartheta_{\textrm{B}}}\left(e^{-i\left(\vartheta_{\textrm{A}}-\vartheta_{\textrm{B}}\right)}+1\right)\left|A_{-}, B_{-}\right>\Bigr\},
\end{split}\label{eq:Output_Interfere_State}
\end{equation}
where $\vartheta_{\textrm{A}}$ and $\vartheta_{\textrm{B}}$ are the phase differences between the laser beams forming the two Bragg splitters, and $\left|A_{+}\right>$, $\left|A_{-}\right>$, $\left|B_{+}\right>$, and $\left|B_{-}\right>$ are the four output modes. Notice that this output state describes genuine inter-modal entanglement, and that all the detection probabilities are directly deducible from the expansion coefficients.

An analysis similar to the one before gives the following expression for the correlation coefficient:
\begin{equation}
\begin{split}
E\left(\vartheta_{\textrm{A}},\vartheta_{\textrm{B}}\right) &= P\left(A_{+},B_{+}\right) + P\left(A_{-},B_{-}\right)\\ 
&- P\left(A_{+},B_{-}\right) - P\left(A_{-},B_{+}\right)\\
&= \cos \left(\vartheta_{\textrm{A}}-\vartheta_{\textrm{B}}\right) = \cos 2\vartheta,
\end{split}\label{eq:corr_four_mode}
\end{equation}
where $P\left(A_{\pm},B_{\pm}\right)$ are the two-atom, joint, detection probabilities. The entropies of entanglement of $\left|\chi\right>$ and $\left|\widetilde{\chi}\right>$ are independent of the relative interferometric output phase, $\vartheta$, as such phases are relative phases in real space. These results are shown in Fig. \ref{fig:f5}(d). In essence, the problems of mode basis transformation—pertaining to quantifying the degree of entanglement—and of mutually commuting components of linear momentum along multiple axes—pertaining to quantifying the Bell nonlocality—are solved by an interferometric arrangement with tunable relative phases.

\section{A Physical Interpretation of the Unitarily Transformed State}\label{sec:Interpret}

Remarkably, the state in Eq. (\ref{eq:Entangled_EPR_State}), $\Psi\left(\nu_{\textrm{A}}, \nu_{\textrm{B}}\right)$ provides a direct and \emph{manifestly nonlocal} description of the entangled particle pair throughout the flight of the photons. I am assuming that this description is based on, for example, knowledge of successful initial state preparation. In contrast, in the absence of such knowledge, one could interpret the state in Eq. (\ref{eq:Transformed_Entangled_State}), $\widetilde{\Psi}\left(\nu_{\textrm{A}}, \nu_{\textrm{B}}\right)$ as encoding reconstructed descriptions of quantum events—or assignments of state vectors—after the quantum measurements have been made; notably, similar ideas have been proposed to explain the observation of delayed-choice entanglement swapping \cite{peres2000delayed, ma2012experimental} in which photons become entangled, after they have already been registered.

More specifically—for the case of photon-photon entanglement—the observers at the two measurement stations, A and B [see, for example, Fig. \ref{fig:f5}(a)] would assign—after performing local measurements on their individual quantum subsystems, and subsequently comparing their expectation catalogs of single and joint detection events—such \emph{ex post facto}, or \emph{after-the-event} quantum state assignments to describe their observations. Equation \ref{eq:Probabilities} gives the probabilities of all such single and joint detection events.

To properly interpret $\widetilde{\Psi}\left(\nu_{\textrm{A}}, \nu_{\textrm{B}}\right)$, one should regard the individual, recorded events—for instance, measurements that reveal probability amplitudes, as well as relative phases at the two space-like separated, measurement stations—as more fundamental than the quantum state, itself. In this viewpoint, therefore, the wavefunction is the description that the observers assign to the overall situation or the entire phenomenon—as it were, after all the events have occurred—and is determined by the complete set of all possible experimental arrangements; for example, of all allowable relative orientations of $\hat{a}$ and $\hat{b}$ in Fig. \ref{fig:f5}(a). Of note, this conception of $\widetilde{\Psi}\left(\nu_{\textrm{A}}, \nu_{\textrm{B}}\right)$ is in harmony with several aspects of the Copenhagen interpretation of quantum mechanics \cite{sep-qm-copenhagen}, especially, as emphasized by Zeilinger and co-authors in the past few decades \cite{RevModPhys.71.S288, Zeilinger1999-tk, bertlmann2013quantum, bertlmann2013quantum2}. The interpretation of the results of a recent elegant experiment, clarifying the subjective nature of path information, specifically, the \emph{which-way information} in three-particle
interferometry, also suggests the primacy of quantum events \cite{jiang2025subjectivenaturepathinformation}. This path information plays a key role in how spatial, \emph{mode-entangled}, or path-entangled superpositions of photons are described.

As is shown in Fig. \ref{fig:f5}(b), the degrees of entanglement of $\Psi\left(\nu_{\textrm{A}}, \nu_{\textrm{B}}\right)$ and $\widetilde{\Psi}\left(\nu_{\textrm{A}}, \nu_{\textrm{B}}\right)$ are identical, regardless of the relative phase, $\theta = \theta_{\textrm{A}} - \theta_{\textrm{B}}$. This invariance of the value of the entanglement with regard to the relative orientation (or basis) is a characteristic defining property of entangled particles. An interesting consequence of this feature is described below.

Recently, experimental violation of Bell's inequalities due to quantum indistinguishability by path identity, as opposed to quantum entanglement, has been reported \cite{wang2025violationbellinequalityunentangled}. As the authors themselves emphasize, to show this effect, they were actively manipulating the state during its creation, rather than merely measuring the properties of an unaltered entangled state \cite{wang2025violationbellinequalityunentangled}. This manipulation is afforded by their experimental setup. In contrast, the usual setup [see, for example, Fig. \ref{fig:f5}(a)] does not change $\widetilde{\Psi}\left(\nu_{\textrm{A}}, \nu_{\textrm{B}}\right)$ while it varies $\theta$ for the purposes of demonstrating the violation of Bell's inequalities. Therefore, in the usual case, the unchanging entangled state, $\widetilde{\Psi}\left(\nu_{\textrm{A}}, \nu_{\textrm{B}}\right)$, or, equivalently, $\Psi\left(\nu_{\textrm{A}}, \nu_{\textrm{B}}\right)$ is responsible for the violation of Bell's inequalities, and the origin of Bell nonlocality.

\section{A More Formal Justification for the Equivalence}\label{sec:Derive}

It is well known that the modes, in quantum mechanics, can be mathematically treated as quantum states. In our approach, we model the spatial optical modes, just like any other bosonic mode, as single-particle wavefunctions. Specifically, we map such modes onto the single-quantum, energy eigenmodes of a simple harmonic oscillator. For example, to model the key term of the input state [see Eq. (\ref{eq:Initial})], that is, $\widetilde {\left|\Psi\right>_{i}} = \left|1\right>_{\textrm{a}}\left|1\right>_{\textrm{b}} \equiv \left|a\right>_{\textrm{1}}\left|b\right>_{\textrm{2}},$ which corresponds to two distinguishable photons occupying two distinct modes, we make the following mappings:
\begin{equation}
\begin{split}
\left|a\right>_{\textrm{1}} &\mapsto \chi\left(x\right),\\
\left|b\right>_{\textrm{2}} &\mapsto \phi\left(y\right),\\
\end{split}\label{eq:input_mappings}
\end{equation}
where $\chi\left(x\right)$ and $\phi\left(y\right)$ are the single-particle, harmonic oscillator energy eigenfunctions, and $x$ and $y$ represent the coordinates of the two particles, respectively. Therefore, we can represent this key term as: $\left|a\right>_{\textrm{1}}\left|b\right>_{\textrm{2}} \mapsto \chi\left(x\right)\phi\left(y\right)$. Strictly speaking, one could model the two optical modes more rigorously as energy eigenmodes of two harmonic oscillators in two distinct Hilbert spaces, respectively; however, one would arrive at exactly the same result. For simplicity, and without loss of generality, I am assuming that we have a single harmonic potential in real space. Let us also consider a global and time-independent anharmonic perturbation that can be applied globally, and that alters the single-particle wavefunctions, from $\chi\left(x\right)$ and $\phi\left(y\right)$, to $\chi^{\prime}\left(x\right)$ and $\phi^{\prime}\left(y\right)$, respectively in a smooth and continuous fashion. As is well known, such energy eigenfunctions of the anharmonically perturbed potential exist, and can be calculated using standard, time-independent perturbation methods.

Figure \ref{fig:f3} describes a simple mechanism that involves making a quantum measurement on an ancillary photon, and that uses a quantum mechanical random process to control the degree of anharmonicity of the potential. Since the successful photo-detection of the incident photon, which is propagating through the Young's double-slit apparatus and is interfering with itself, is a quantum mechanically probabilistic process, the state of the entire photo-detecting apparatus is brought into a macroscopic superposition of having detected and having not detected the photon. Consequently, the initial state, $\chi\left(x\right)\phi\left(y\right)$ is transformed into this entangled superposition: $\gamma \chi\left(x\right)\phi\left(y\right) + \delta \chi^{\prime}\left(x\right)\phi^{\prime}\left(y\right)$, where $\gamma$ and $\delta$ are suitable normalizing, expansion coefficients.

Now, we can make these inverse mappings to recover the optical modes from the corresponding energy eigenfunctions of the anharmonically perturbed potential:
\begin{equation}
\begin{split}
\chi^{\prime}\left(x\right) &\mapsto \left|a^{\prime}\right>_{\textrm{1}},\\
\phi^{\prime}\left(y\right) &\mapsto \left|b^{\prime}\right>_{\textrm{2}},\\
\end{split}\label{eq:output_mappings}
\end{equation}
where the spatial coordinates, $x$ and $y$, and the particle labels, $1$ and $2$ have the same meanings, as before. Notice that the above step is the converse of the one shown in Eq. (\ref{eq:input_mappings}), and assumes that these mappings are invertible, which is usually the case. For simplicity, and to avoid any inadvertent pathologies, I am going to assume that this mapping is one-to-one and onto.

Therefore, the initial state, $\widetilde {\left|\Psi\right>_{i}}$ has been transformed to:
\begin{equation}
\begin{split}
\widetilde {\left|\Psi\right>_{f}} = \gamma \left|a\right>_1\left|b\right>_2 + \left|a^{\prime}\right>_1\left|b^{\prime}\right>_2,\\
\end{split}\label{eq:final_state}
\end{equation}
which is a genuinely entangled state of two interacting particles. Of note, as is evident from this simple argument, creating such a state requires access to at least four distinct spatial optical modes, and two incident and initially unentangled optical photons. Notice that this transformation process is not completely deterministic, as it is conditioned on the probabilistic photo-detection of the ancillary, trigger photon 3.

\bibliography{Bibliography}

\end{document}